\documentclass[10pt,a4paper]{article}

\usepackage[utf8]{inputenc}
\usepackage[T1]{fontenc}
\usepackage{microtype}
\usepackage[sc]{mathpazo}
\usepackage{graphicx}

\usepackage[bottom, hang]{footmisc}
\usepackage[
    top=2.5cm,
    bottom=2.5cm,
    left=2cm,
    right=2cm,
    footskip=1cm,
    headsep=0.75cm,
    columnsep=20pt,
]{geometry}

\usepackage{fancyhdr}
\patchcmd{\maketitle}{plain}{empty}{}{}

\renewenvironment{abstract}{
  \begin{center}
    \normalfont\bfseries\vspace{0.5\baselineskip} \abstractname
  \end{center}
  \begin{quote}
  \normalfont\small\itshape
}{\end{quote}}

\usepackage{natbib}
\setcitestyle{authoryear}
\usepackage{titling}
\pretitle{\begin{center}\huge\bfseries}
\posttitle{\end{center}}
\patchcmd{\maketitle}{plain}{empty}{}{}

\usepackage{booktabs}

\usepackage{caption}
\usepackage{enumitem}
\setlist{noitemsep}

\usepackage{hyperref}

\usepackage{amsmath,amssymb}

\title{Card Sorting Simulator: Augmenting Design of Logical Information Architectures with Large Language Models}

\author{Eduard Kuric\textsuperscript{1,2,}\thanks{Corresponding author: \href{mailto:eduard.kuric@stuba.sk}{eduard.kuric@stuba.sk}\\ORCID(s): 0000-0002-7371-5512 (E. Kuric), 0000-0002-4111-1052 (P. Demcak), 0000-0001-9030-7337 (M. Krajcovic)\\ \textbf{Citation:} Kuric E., Demcak P., Krajcovic M. (2025) Card Sorting Simulator: Augmenting Design of Logical Information Architectures with Large Language Models, https://doi.org/XXXXXXX.XXXXXXX\\ \copyright2025 Copyright for this paper by its authors. Use permitted under Creative Commons License Attribution 4.0 International (CC BY 4.0)} , Peter Demcak\textsuperscript{1} and Matus Krajcovic\textsuperscript{2,1}
}

\date{
\footnotesize\textsuperscript{1}
Faculty of Informatics and Information Technologies, Slovak University of Technology, Ilkovicova 2, Bratislava, 84216, Slovakia\\ \textsuperscript{2}
UXtweak Research, Cajakova 18, Bratislava, 81105, Slovakia\\
}

\begin{document}

\maketitle

\begin{abstract}
{\textit
Card sorting is a common ideation technique that elicits information on users' mental organization of content and functionality by having them sort items into categories. For more robust card sorting research, digital card sorting tools could benefit from providing quick automated feedback. Our objective of this research is to advance toward an instrument that applies artificial intelligence (AI) to augment card sorting. For this purpose, we develop the Card Sorting Simulator, a prototype tool that leverages Large Language Models (LLMs) to generate informative categorizations of cards. To illuminate how aligned the simulation is with card sorting by actual participants, and to inform the instrument’s design decisions, we conducted a generalizability-focused comparative study. We obtained 28 pre-existing card sorting studies from real practitioners, comprising 1,399 participants, along with diverse contents and origins. With this dataset, we conducted a comprehensive and nuanced analysis of the agreement between actual card sorting results (clusterings of cards) and synthetic clusterings across a multitude of LLMs and prompt designs. Mutual information scores indicate a good degree of agreement to real result clustering, although similarity matrices also demonstrate inconsistencies from mental models, which can be attributed to their top-down nature. Furthermore, the number of cards or complexity of their labels impact the accuracy of its simulation. These findings bolster the case for AI augmentation in card sorting research as a source of meaningful preliminary feedback and highlight the need for further study for the development and validation of intelligent user research tools.}
\end{abstract}

\begin{center}
\end{center}

\section{Introduction}

Today’s digital solutions tend to present labyrinthine amounts of functionality and content, making them difficult to navigate without proper organization. To construct intuitive information architectures—menus, structural frameworks—user experience (UX) professionals often rely on the well-established method of card sorting to uncover the mental models by which users naturally group information \citep{tankala2024, baxter2015, lewis2021, munim2023, paea2022, katsanos2023, katsanos2022, best2021, jiang2021a, macias2021, naranjorojas2022, fauth2024, feine2019}. Nonetheless, qualitative research consumes significant amounts of time and labor, delaying its outcomes \citep{gerosa2024}. This runs counter to design processes that thrive on rapid iteration and timely decision-making, such as by evaluating low-fidelity prototypes early, even at some cost to accuracy \citep{krajcovic2025, ishaq2021}. What if we could redefine what is possible by delivering quick feedback on labels and logic even as card sorting is still at the drawing board, maximizing the value received from participant responses from the start? Could we reduce the likelihood—or the necessity—for reiterating card sorting exercises to gain additional insights while remaining mindful of project budget constraints?

Guided by principles of human-centered artificial intelligence (AI), we introduce an instrument for augmenting the capabilities of UX researchers in planning, conducting and analyzing information architecture research. Using Large Language Models (LLMs), we implement the Card Sorting Simulator, a foundational subsystem for exploring logical item groupings of items as approximations of human mental models. To date, LLM-simulated card sorting has not been thoroughly investigated for similarities to groupings perceived as logical by actual users. Their alignment could be further influenced by a range of factors including domain, audience demographics, knowledge and habits, prompt design, model selection (e.g., GPT, Gemini, Claude, DeepSeek), label phrasing, complexity, number of cards, participant instructions, and the broader context. Therefore, to validate our instrument and identify effective choices for its design, we conducted a large-scale comparative analysis across 28 diverse pre-existing real-world card sorting studies encompassing 1,399 participants.

Our results indicate that (1) LLM-driven card sorting can predict the most prominent patterns from human participants to a reasonable degree, although (2) there can be disagreement about the placement of individual cards, conceptual levels and perspectives of created categories and a lack of realistic diversity. Meanwhile (3) the accuracy is modulated by the complexity of the card sorting exercise. Paradoxically, (4) a high-level prompting strategy of generating aggregate clusterings can produce more accurate results than low-level simulation of individual participants. For optimization of the prototype instrument, (5) LLM model selection does not have noteworthy impact, nor is it significantly enhanced by incorporating context that is implicitly present in card sorting studies (e.g., custom instructions, recruitment criteria).

This research offers both practical and theoretical contributions. On the practical level, the proposed instrument incorporates a prototype of card sorting simulation to streamline the evaluation of information architectures by incorporating low-fidelity, rapid AI feedback on top of high-fidelity, effort-intensive human feedback. This combined approach can serve as the foundation for designing information architectures that are more intuitively navigable by users. From a theoretical standpoint, our empirical investigation provides in-depth insight into the ability of LLMs to logically group concepts. Our findings underpin card sorting simulation as an augmented supplement to, rather than a replacement for, traditional card sorting, highlighting the critical role of the human factor in UX research.

The structure of this article includes the following sections: Section \ref{sec:2} analyzes the background of card sorting and its augmentation, Section \ref{sec:3} introduces our instrument for AI-augmented card sorting, Section \ref{sec:4} presents its evaluation through a rigorous experiment methodology and Section \ref{sec:5} analyzes the results of said experiment. In Section \ref{sec:6}, the data is further discussed to establish the findings, their implications and limitations, and to suggest prospective future work. Finally, Section \ref{sec:7} represents the conclusion.

\section{Related work and background}
\label{sec:2}

Card sorting is a common and easy-to-use method that harnesses users’ mental organization of concepts to support the design of information architectures that make sense to them \citep{macias2021}. This section provides an overview of knowledge relevant to its automation and augmentation.

\subsection{Card sorting user research method}
\label{sec:cardsort}

Card sorting is widely used for organizing and optimizing information architectures \citep{best2021, jiang2021a, macias2021, naranjorojas2022, fauth2024}. It tasks participants with categorizing labeled cards—the information items in need of organizing—based on their perceived connections and logical groupings \citep{rosenfeld2015, schall2014}. This approach helps uncover how users naturally structure information, ensuring that the content labeling and organization align with their mental models \citep{tankala2024}. 

Implementing card sorting can lead to more intuitive navigation and improved usability \citep{ntouvaleti2022}. As such, it is applied to improve information architectures across various fields and industries, including government \citep{jiang2021a}, insurance \citep{thomas2013}, libraries \citep{lewis2009} and healthcare \citep{wentzel2016, munim2023}. Mismatches between naive taxonomies created by users during card sorting and information structures used for navigation do not inherently reduce the navigation task performance \citep{schmettow2016}. Therefore, to validate an information architecture, card sorting is commonly followed by tree testing \citep{kuric2025treetest}

Card sorting has three main types \citep{robles2019, greifeneder2022}:
\begin{itemize}
    \item \textbf{Open card sorting.} Participants sort cards into categories of their own making. This produces diverse and natural representations of users’ mental models, since no pair of individuals creates categories that are exactly the same.
    \item \textbf{Closed card sorting.} Users sort cards into a predefined taxonomy of categories to assess its alignment with the mental models of users. However, this approach prevents users from expressing their own classifications
    \item \textbf{Hybrid card sorting.} A combination of open and closed sorting, allowing users to make use of predefined categories, or discard them in favor of original ones.
\end{itemize}

While the card sorting process can be flexible in some methodological aspects \citep{tankala2024, tchivi2025}, it can usually be logically divided into three phases. Remote online card sorting tools are also typically structured to accommodate this flow:
\begin{itemize}
    \item \textbf{Planning and Configuration.} UX researchers decide upon the type of card sorting, the number of cards and create card labels. Optional details contained in the configuration may also include messaging (e.g., instructions) and questionnaires (e.g., screening or profiling of participants). Configuration should avoid introducing biases into categorization.
    \item \textbf{Conducting and Data Collection.} Participants are recruited and card sorting sessions are held either in person or remotely. The sessions can be moderated (for deeper insights) or unmoderated (for larger samples).
    \item \textbf{Analysis of Results.} Data obtained during the previous step is subjected to analysis, extracting patterns from sorting results through statistical and clustering techniques.
\end{itemize}

Upon completion of data collection, a variety of data analysis techniques can be employed. The preference for them first depends on the card sorting variant. Since open card sorting does not involve a unifying preset of categories, its results are typically converted into aggregations that are suitable for further analysis. The similarity matrix is one such aggregation, consisting of a grid with cards placed along its axes \citep{lantz2019, paea2022, bussolon2008, martin2023}. Cell values represent how often each pair of cards was grouped together, regardless of the individual categories.

Similarity matrices can be visualized as heatmaps, or serve as the input for clustering algorithms. Two of the most common approaches include dissimilarity analysis and multivariate analysis. Dissimilarity analysis involves dendrogram techniques such as the Best Merge Method and Wards’ Method that rely on pairwise measures of distance \citep{paea2022, ali2019, righi2013}. Multivariate analysis algorithms such as Principal Component Analysis (PCA) and Multidimensional Scaling (MDS) can be used to reduce the dimensionality of raw distances, and k-means can be applied for clustering \citep{macias2021, gabethomas2016}. 

In contrast, closed card sorting studies lend themselves more to a descriptive and straightforward analysis aimed at validation of the category preset rather than exploration \citep{robertson2021}. As such, their characteristic indicators are more category-based. Examples can include the intra-category level agreement, cards commonly placed within a category, agreement about the placement of individual cards, and the order of categories into which an individual card was placed.

\subsection{AI-augmented user research}

Augmentation of UX research methodologies with AI technologies is an expanding area of study, yet card sorting is lagging behind as underexplored compared other methods in terms of AI-driven assistance and semi-automation \citep{lu2024b}. Its outcome can be a “symbiosis” where AI enhances the understanding capabilities of researchers while the performance of AI is also improved \citep{raees2024}. These endeavors can be attributed to the rising prominence of human-centered AI, a paradigm that places humans at the forefront of decision making. Instead of attempting to substitute humans with simulacra, AI methods are regarded as instruments to fulfill human needs, which should be explainable and guided by ethical principles \citep{capel2023}.

The strengths of AI for processing and extracting patterns from vast amounts of data are well-suited for analysis of UX research data. Typical use cases include the generation of transcripts, summaries and labels from user interviews \citep{aitim2024, borlinghaus2021, jiang2021b}, or the analysis of user reviews to extract sentiment, satisfaction and other types of feedback \citep{guzman2014, jang2022}. More recent advancements in Large Language Models (LLMs) have also led to the development of new methods for research ideation, such as generation of user personas \citep{schuller2024}

Promisingly, LLM chatbots, with their strong communicative abilities \citep{gessinger2025}, could boost the interactiveness of UX research by facilitating adaptive conversations with participants in the natural language. Conversational surveys can drive participant engagement and enhance the quality of harnessed information \citep{xiao2020}. Investigation of the ability of LLMs to automatically ask reasonable follow-up questions after tasks in unmoderated usability testing revealed a number of challenges \citep{kuric2024gpt}. Although the answers did provide more details, they did not identify more usability issues and contributed to heightened frustration and repetition. 

The research note by \citet{rothschild2024} on LLMs in surveys supports the notion of LLMs as potentially transformative for card sorting, given their impact on a related, exploratory and user-centered method. In surveys, LLMs could augment the question design with quick feedback, assist with the refinement of questions, facilitate data cleaning, or power new analytical tools. Simulated user personas could be used to generate synthetic responses. However, scrutiny toward AI-generated responses was stressed, emphasizing the priority of their empirical validation. Promising applications of AI translate into a number of risks, including the lack of transparency, limited representativeness, propagation of biases and stereotypes, the loss of authenticity and researcher control. Hybrid methods that espouse the cooperation of human researchers and participants with AI could represent viable solutions to these issues \citep{gerosa2024}.

\subsection{Synthetic results in card sorting}

Simulation of card sorting is yet to be thoroughly and systematically investigated. Past attempts at automation primarily concerned the mass collection of data from humans \citep{zavod2002}. This is natural, given the focus of card sorting on assessing mental models and taxonomies of real people. At most, prediction was used as an analytical tool, with decision trees to predict how individual cards were sorted based on the profiles of existing participants \citep{martin2023}. 

To the best of our knowledge, the sole investigation of an LLM synthesis of card sorting results was conducted by \citet{sauro2024}. ChatGPT’s categorization of 40 ecommerce products was compared with the aggregated sorting by 200 participants. Some similarities were observed in the number of created categories (5) and an average 68\% match in the placement of cards within corresponding categories. This constitutes preliminary evidence that LLM-generated card sorting could be valuable, even if it does not match the depth of human judgement. The study’s focus on a single card sort limits its generalizability. Its presumption of straightforward compatibility between real and synthetic categories in a one-to-one relationship compels a more in-depth and robust evaluation.

The feasibility of simulated card sorting could be supported through a broader lens of synthetic generation of results in UX research. This area of study faces similar challenges with potentially revealing implications. LLMs have been explored in the fields of UX research and design for their ability to simulate human responses. The descriptions of previous experiences provided by users and LLMs are similar enough to be challenging for humans to distinguish effectively \citep{hamalainen2023, hamalainen2022}. Idea formation and pilot evaluation represent two practical applications for quick artificial responses. As a drawback however, they also have the potential to cause harm in online crowdsourcing platforms by facilitating the fabrication of fake responses. \citet{gu2024} introduced a chatbot for assisting designers with familiarizing themselves with user personas, which did not achieve significant improvements over static personas. \citet{gerosa2024} suggested three levels of synthesis based on the number of persons being simulated by the approach: single- (a user), multi- (a focus group) and mega-persona (a population).

In usability assessment, LLMs were also leveraged for discovery of usability issues, or to provide UX research assistance \citep{kuric2025slr}. SimUser \citep{xiang2024} is a tool that simulates the perception and interaction between users and a mobile application through the conversation of two agents with chain-of-thought reasoning. While capable of uncovering a percentage of usability issues, its feedback was different from information expressed by participants. Examples include placing emphasis on aspects that no participant focused on, tolerance toward frustrating experiences, the inability to draw upon specific expectations and experiences of genuine users, or to reflect their characteristics.

In social research, LLMs have also been central to the design of new methods aimed at simulating answers to surveys and polls. \citet{kim2024} introduced a framework for opinion prediction in questions that either weren’t asked during a specific year, or were not included in the training data. Their findings suggested LLMs are best fit as a supplement of surveys, particularly for imputing missing data, where the accuracy improves as more questions are used for fine-tuning. \citet{sanders2023} developed a method for generating human-like responses about policy using GPT, showing strong correlation with real polling data but also deficient ability to extrapolate to recent world events.

To summarize the current landscape, card sorting, a traditional user research technique, could be enhanced by integrating AI-based methods to assist researchers, including the simulation of card sorting results as a supplementary source of feedback. However, the gap addressed by this research stems from minimal validation of such approaches, as well as a lack of established methods or tools.

\section{Instrument design}
\label{sec:3}

The main goal of this research is to design and validate the prototype of a system that facilitates AI-augmented card sorting. By nature, this extends card sorting as a well-established formative and exploratory user experience research method \citep{paea2022, tchivi2025}. In standard card sorting, mental models are inferred directly from users based on how they organize content. However, researchers also rely on their own expertise to plan, configure, and analyze card sorting, which can limit the scale and depth of the perspectives being explored, as well as introduce various potential biases. 

Therefore, we propose an instrument that can assist in decision-making processes by supporting researchers with quick actionable insights, such as providing feedback on card selection and presentation, and facilitating changes before recruitment of actual participants. This section describes the instrument, as well as its focal component, the LLM-driven Card Sorting Simulator, which we implemented and evaluated to assess the conceptual viability of the instrument.

\subsection{AI augmentation in card sorting process}

As the initial step, we identified scenarios that could benefit from positive transformation through the introduction of semi-automated card sorting. Instead of seeking to simply substitute current manual procedures with automation or fit automation into methodological frameworks that evolved to function without automation, we adopted a perspective aimed at addressing underlying needs, so that methodology can be revised in the light of these considerations. Specifically, we considered aspects that can hinder iteration or make card sorting cumbersome. Consequently, our design tackles the following barriers: (1) unexpected impact of design variables, (2) susceptibility to changes, (3) re-recruitment, and (4) gap between ideation and validation.

\textit{Unexpected impact of design variables.} Even minor differences in the choice of cards, card labels or recruitment criteria can significantly impact the resulting categorization of concepts. Therefore, during the stage of initial planning, AI augmentation could offer assistance by generating preliminary feedback and support troubleshooting. For example, the quick surface-level System 1 reasoning of a Large Language Models \citep{li2025} may reveal logical shortcuts that participants could take in their own sorting, although researchers would prefer them to consider the cards in greater depth. 

\textit{Susceptibility to changes.} Results from card sorting are linked inextricably to a specific set of cards and conditions. For various reasons, researchers may need to assess how users would sort cards under slightly altered conditions. For example, this can occur after incorporating participant feedback, to change the type of card sorting (open, closed, hybrid), or when there is too much content and functionality to sort during a single session. However, limited time or budget may prevent researchers from conducting further iterations. Comparison of obtained data to AI-generated sorting with an adjusted list of cards could enhance the ability of researchers to understand how their changes could impact mental processing of concepts or obtain evidence for further iterations being necessary.

\textit{Re-recruitment.} The sample that is recruited for card sorting may not sufficiently cover all relevant types of users. Participant recruitment is among the most costly aspects of UX research, which means that its scale can be limited. AI can capture characteristics of certain groups of users \citep{xiang2024}. This could be applied for heuristic assessment of potential similarities and differences in sorting among groups that are underrepresented in the available sample. User personas that reflect inclusion/exclusion criteria during recruitment can be regarded as key inputs for simulated sorting. 

\textit{Gap between ideation and validation.} Aside from card sorting for idea generation, creating a user-centered information architecture also requires tree testing to validate its design \citep{kuric2025treetest}. The need to conduct and analyze each study separately, along with the potential need for multiple iterations, can make the process cumbersome and time-consuming. AI assistance capable of automatically sorting cards could be implemented to support a rapid version of tree testing. When users encounter difficulties finding information within the tree, the architecture could adapt reactively based on logical categories identified by an AI algorithm. The next batch of participants could then be used to assess the impact of changes on task performance. The information architecture is thus optimized semi-automatically, reducing delays between iterations.

\begin{figure}[!ht]
    \centering
    \includegraphics[width=0.7\linewidth]{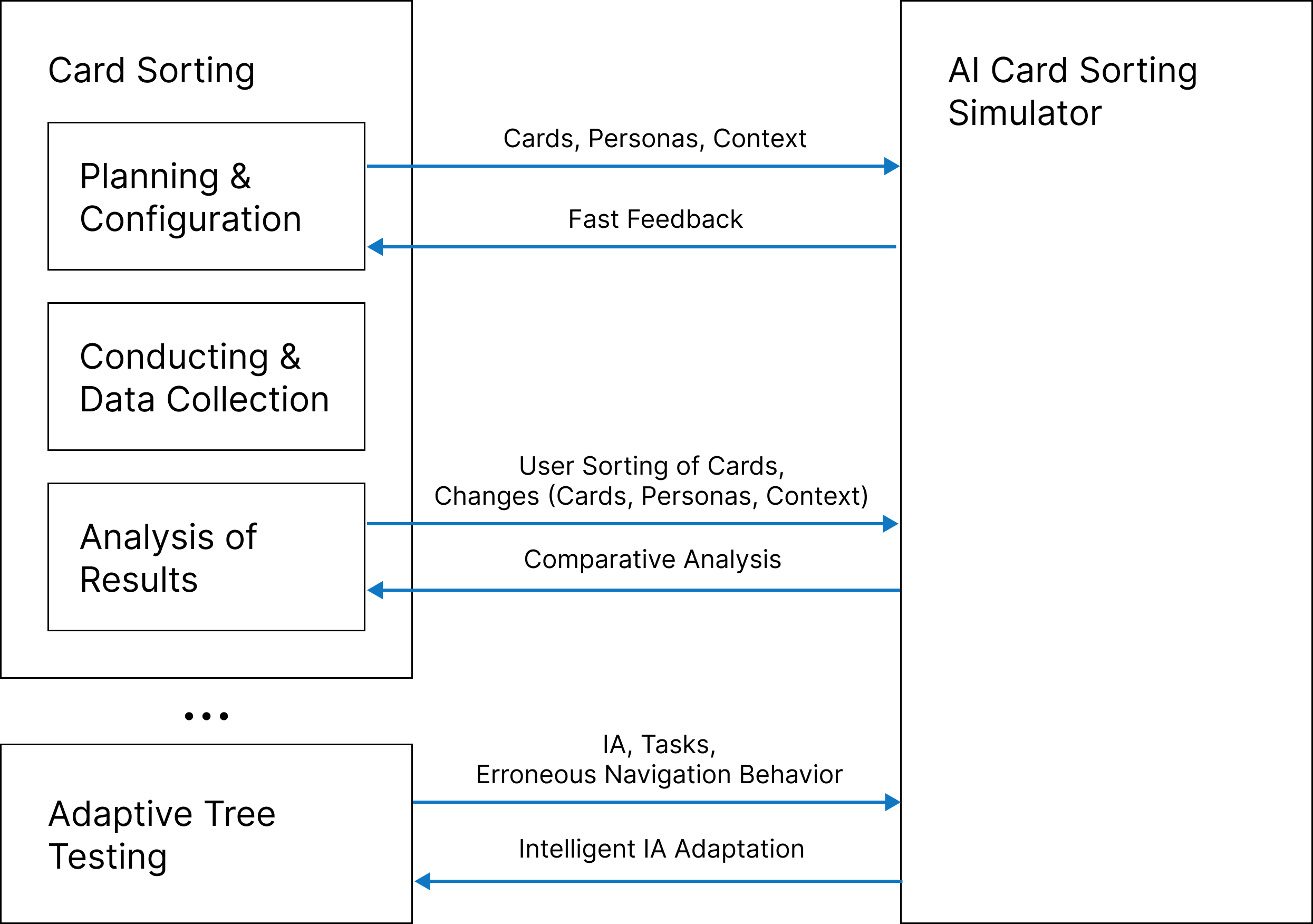}
    \caption{Overview of AI-augumented Information Architecture (IA) ideation and validation workflow. Based on card sorting and tree testing information, as well as user (UX researcher) analytical demands, the Card Sorting Simulator generates outputs that support interactive analysis and rapid adaptation to findings.}
    \label{fig:ia-process}
\end{figure}

The proposed system enables a workflow (see \autoref{fig:ia-process}) that enhances and streamlines card sorting and tree testing processes in UX research tools and environments. Information from card sorting and tree testing methods submitted by researchers (e.g., card labels, participant personas, instructions) serves as the input to the AI-driven Card Sorting Simulator module, with the goal of sorting cards in a manner approximating mental models of human users. By applying further heuristic and analytical methods to process and interpret AI-generated sorting results, the system traverses the scale between augmentation and methods that incorporate more comprehensively automated elements \citep{esposito2024} (e.g., highlighting cards that are likely to be alone in their category vs. tree testing that adjusts IA mid-evaluation to automatically assess mental model variants).

\subsection{General design decisions}

The prerequisite for implementing the system for AI-augmented research of information architectures lies in the availability of an AI model that can solve card sorting tasks in a manner that reflects solutions by real users, with a reasonable degree of accuracy in its representativeness of the intended user population. Therefore, the subject of this research focuses on the design of a prototype of the AI component of the Card Sorting Simulator module and the evaluation of its outputs.

Due to the experimental nature of the designed system, several salient design decisions were considered and implemented as parallel variants of the simulator. By thorough comparative evaluation, the optimal version can be selected and assessed. Additionally, in the case of significant differences between conditions, guidelines for optimizing the selection of the model’s parameters based on the situational context can be formed. The following design decisions were made:
\begin{itemize}
    \item \textit{LLM model choice.} Four state-of-the-art Large Language Models were chosen: GPT-4o\footnote{ChatGPT: \url{https://chatgpt.com/}} (OpenAI) \citep{openai2024}, Gemini\footnote{Gemini: \url{https://gemini.google.com/}} 2.0 Pro (Google) \citep{google2023}, Claude\footnote{Claude: \url{https://www.anthropic.com/claude}} 3.5 Sonnet (Anthropic) and DeepSeek\footnote{Deepseek: \url{https://www.deepseek.com/}} V3 (latest available versions in February 2025). All assessed models provide advanced text processing capabilities, which can be regarded as a prerequisite for simulation of human card sorting tendencies \citep{lu2024, lippert2024, wang2023}. As a baseline, they were configured with default parameters (e.g., temperature) to mirror their configuration in web-based chatbots.
    \item \textit{Prompt and output format.} Large Language Models are black box models. As such, even minor changes to the structure and contents of prompts can have unpredictable and significant impacts on their output. Therefore, prompt format is an essential factor in designing an LLM-driven system \citep{chang2024}. To increase the generalizability of findings made in the evaluation of the Card Sorting Simulator, we implement multiple logical conceptions for prompting LLM models and processing their outputs (see \ref{sec:prototype} \hyperref[sec:prototype]{Prototype technical implementation} for their specification).
    \item \textit{Cart sorting variant.} To balance scope and focus in our research, we restrict the designed simulation to the open variant of card sorting (see \ref{sec:cardsort} \hyperref[sec:cardsort]{Card sorting user research method}). The open variant was chosen as the most cognitively complex and variable type of the method, given its lack of a predefined list of categories that the participants—or an AI model—can use as a template.
\end{itemize}

\subsection{Prototype technical implementation}
\label{sec:prototype}

LLM prompts for the Card Sorting Simulator were refined by applying multiple prompt engineering techniques aimed at improving their outputs \citep{marvin2024, white2023}. To facilitate their seamless reapplication across different studies of open card sorting, prompts were created as templates. Patterns incorporated in the templates include clear structure, clarification of context \citep{marvin2024} and information about personas that the model is asked to portray \citep{xiang2024, white2023} in the form of sample demographics (e.g., gender and age distributions, occupations).

\begin{figure}[!ht]
    \centering
    \includegraphics[width=\linewidth]{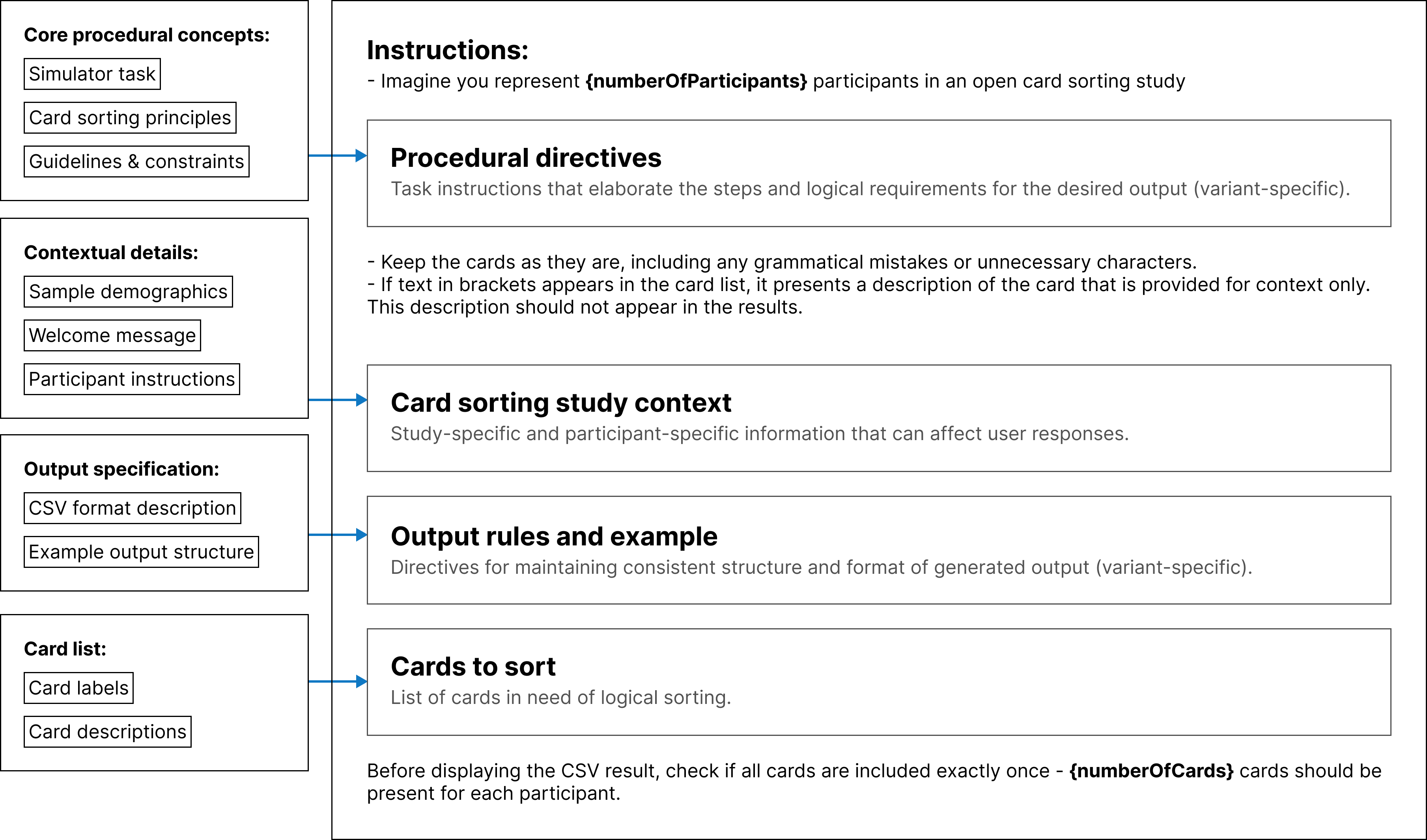}
    \caption{General structure of the prompt template used for generating card sorting results. Sections are clearly delimited. Input information (left) originates from the configuration of the card sorting study (contextual details, card list) or is defined as part of the approach for generating results (experimental design variant). For specific prompt variant templates, refer to \autoref{app:templates} \hyperref[app:templates]{Prompt templates}.}
    \label{fig:prompt-structure}
\end{figure}

Prompts were designed through an iterative process of continuous empirical evaluation in the context of multiple card sorting studies and LLM models. The goal of this process was to achieve specificity and clarity, mitigate errors (e.g., missing cards), as well as biases caused by incorrect interpretation of its fundamental directives. The majority of issues encountered during development were observed in studies with high numbers of cards. The general template of the prompt designs is shown in \autoref{fig:prompt-structure}.

A directive that prompts an LLM to perform card sorting with a given list of items should clearly specify the output format to ensure the consistency of data presentation and processing procedures. Furthermore, the output format, could influence the characteristics of the output content. Therefore, to ensure consistency and assess the impact of the output format, we identified and developed four experimental prompt design variants (P1-P4):
\begin{itemize}
    \item \textit{P1: Raw Data Simulation.} This variant is designed as a faithful artificial recreation of the card sorting activity with actual users. Standard card sorting data is obtained individual people. To simulate this, the prompt instructs the model to generate raw card sorting data while adopting the personas of individual users.
    \item \textit{P2: Similarity Matrix Generation.} Abstracting away from the simulation of individual users, the LLM is tasked with generating a similarity matrix. Similarity matrices are common quantitative representations of open card sorting results, summarizing the number of times that each pair of cards was categorized together as symbols of their perceived similarity.
    \item \textit{P3: Clustering Generation.} The LLM is tasked with creating a definitive system of categories, where each card is placed into a single category based on which category fits it best according to the aggregate results of card sorting from a sample of a population.
    \item \textit{P4: Clustering Generation without Context.} Identical to P3, but without any contextual information (e.g., participant instructions, targeting) to assess its impact.
\end{itemize}

All prompt variants follow the general prompt template in \autoref{fig:prompt-structure}, with differences only in the contents of the procedural directives, output rules and examples to correspond with the underlying differences in their principles. Additionally, P4 does not include the card sorting study context section. An overview of variant-specific prompt templates is available in \autoref{app:templates} \hyperref[app:templates]{Prompt templates}.

\section{Experimental methodology}
\label{sec:4}

For a multidimensional experimental evaluation of the proposed Card Sorting Simulator instrument, we posed several research questions involving its performance and the effects of various design and environmental variables. In this section, we describe the experiment, designed to obtain highly ecologically valid findings through an extensive analysis of a diverse dataset of card sorting studies. These studies were performed in real-world scenarios by third-party UX researchers collecting data for actual decision-making purposes across different fields, organizations and contexts.

\subsection{Research questions}

The general question of our research revolves around the accuracy of LLM-generated synthetic card sorting results. Accuracy of synthetic results is interpreted as their representativeness of the mental models of actual users. Traditional card sorting as an established research method can be considered as the baseline, and its results as the ground truth. Our focus is on identifying key similarities and differences, as well as influences of other factors such as the choice of LLM, prompt design, and study parameters. To investigate these aspects, we define the following research questions and hypotheses:\\

\textbf{\textit{RQ1: Do LLM-generated card sorting results align with those produced by real participants?}}

We anticipate that some degree of similarity will be found in overarching patterns between synthetic and real sorting results \citep{hamalainen2023}, but also that they will exhibit some distinct characteristics. While LLMs can learn logical links between concepts, they are trained on large datasets of text from the internet, whereas users’ mental models can be contextually-specific or influenced by hidden factors. As a null hypothesis, we could predict that the results between the two methods will be identical, which would also favor more holistic automation instead of the proposed augmentation. 

As a benchmark of the simulator's best performance, the best-performing LLM output format is determined. Therefore, prompt designs P1 (raw data), P2 (similarity matrix) and P3 (clustering) are compared in parallel. We expect the accuracy to increase with higher abstraction, given that the task of modeling card sorting at lower logical levels (e.g., simulating individual participants) is more complex. Therefore, our hypotheses state:

\begin{itemize}
    \item \emph{H1a: With clustering generation as its output, LLM-generated card sorting yields results that are most similar to the outcomes of card sorting with real participants.}
    \item \emph{H1b: LLM-generated card sorting results are similar to outcomes of card sorting with real participants, but they do not reach levels of agreement sufficient to serve as direct substitutes.}
\end{itemize}

\textit{\textbf{RQ2: How does the selection of the LLM model affect the similarity of LLM-generated card sorting results to data from real participants?}}

Assessing the differences between the outputs of multiple state-of-the-art LLMs is valuable for optimization of the Card Sorting Simulator. Although various LLM models can behave differently \citep{zhao2023}, we do not anticipate that any of the evaluated models will outperform the others. As generalistic models, neither have been identified to have significant theoretical advantages or disadvantages under consistent conditions. We hypothesize that:

\begin{itemize}
    \item \textit{H2: Across various LLMs, there are no significant differences between the similarity of their card sorting results to the outcomes of card sorting with real participants.}
\end{itemize}

\textbf{\textit{RQ3: How does inclusion of context in the prompt affect the similarity of LLM-generated card sorting results to data from real participants?}}

Context (e.g., target participant demographics) provides the Card Sorting Simulator with additional information that can affect mental models of real users. If the model is capable of suitably adapting to the context, it could boost the accuracy of LLM-generated results. According to ChatGPT documentation, relevant context can be used to generate outputs that adapt to the context \citep{openai2025}. However the capacity of the model to react appropriately is unknown, potentially subject to broader factors on a metacontextual level. We compare conditions where LLM is prompted with context (P3) and without context (P4). We hypothesize that:

\begin{itemize}
    \item \textit{H3: LLM-generated card sorting yields results that are more similar to the outcomes of card sorting with real participants if the prompt includes context.}
\end{itemize}

\textbf{\textit{RQ4: How does the complexity of a card sorting study (number of cards, complexity of card labels) affect its LLM-generated results?}}

More complex card sorting studies are more difficult even for human participants to complete, which is why higher complexity could present barriers for an AI model aiming to approximate them. To better understand the impact of the properties of a card sorting study on its LLM-generated results, we explore the influence of the number of cards and label complexity. We hypothesize:
\begin{itemize}
    \item \textit{H4: The number of cards and label complexity have a negative effect on the similarity of LLM-generated card sorting results to the outcomes of card sorting with real participants.}
\end{itemize}

\textbf{\textit{RQ5: What are the common issues that appear when LLMs are used to simulate card sorting?}}

LLMs in user research have inherent limitations such as a lack of diversity \citep{hamalainen2023}. Aside from questions about the accuracy of simulated results, other issues could arise due to hallucinations, lack of true comprehension, non-determinism, prompt sensitivity, and biases learned from training data or because of the training method. These challenges could present risks for simulated card sorting and its application in augmented UX research processes, unless they are automatically accounted for. We hypothesize that:

\begin{itemize}
    \item \textit{H5: LLM-generated card sorting results are prone to issues that can produce outputs with reduced usefulness, including hallucinations.}
\end{itemize}

\subsection{Experiment design and procedure}
\label{sec:design}

We compare the proposed Card Sorting Simulator to standard card sorting with human participants. In the wild, card sorting can be affected by a variety of variables that could confound our research and reduce the generalizability of our findings (e.g., topic, domain, researcher’s practices and preferences, demographics of participants). Therefore, we placed high priority on ecological validity by conducting a large-scale experiment where our Card Sorting Simulator was applied to a high number of diverse card sorting studies that were conducted by third-party UX professionals in real practice.

UXtweak\footnote{UXtweak UX research tool: \url{https://www.uxtweak.com/}}—a widely adopted online UX research tool applied annually in thousands of card sorting studies—provided the anonymized data necessary for the systematic evaluation. Data from UXtweak’s users was obtained in agreement with UXtweak’s internal ethical and privacy policies, as well as the legal norms of the EU, ensuring compliance with the GDPR and other relevant regulations. Among users who conducted at least one open card sorting study in 2024, a curated group was contacted with information explaining the research and a voluntary consent request. Consenters granted us permission to create a copy of a single randomly chosen open card sorting study from their account.

Data was automatically anonymized before processing to remove participants’ personally identifiable information, if any was submitted in optional questionnaires incorporated in the studies. To mitigate biases during analysis, data quality assessment and descriptive coding of studies was conducted by two researchers, then a third researcher resolved all conflicts that emerged. To ensure that the card sorting studies used to evaluate the Card Sorting Simulator were of sufficient quality to yield relevant findings, their quality was reviewed beforehand. Following inclusion criteria were applied:

\begin{itemize}
    \item At least 10 participants
    \item At least 10 unique cards
    \item No duplicate cards
    \item English language
    \item Standard card sorting and with reasonable contents (to exclude studies that could use the Card Sorting tool for nonstandard purposes)
\end{itemize}

Given that UXtweak Card Sorting is an online research tool where participants can sometimes abandon a study without completing it, some studies also contained data from partially completed responses. However, only complete responses from participants who categorized all of the cards in accordance with the requirements of card sorting were included in the calculation of the results for the purposes of the analysis.

After the review, a set of card sorting data was prepared for analysis. Information about individual card sorting studies was extracted to complete LLM prompt templates and obtain final prompts. Synthetic card sorting results were generated by the LLM. This data was subjected to comparative analysis with the results obtained in the original card sorting studies. See \ref{sec:simulation} \hyperref[sec:simulation]{Card sorting simulation and data processing} for details about execution of the simulations.

\subsection{Dataset}

\begin{figure}[!ht]
    \centering
    \includegraphics[width=0.7\linewidth]{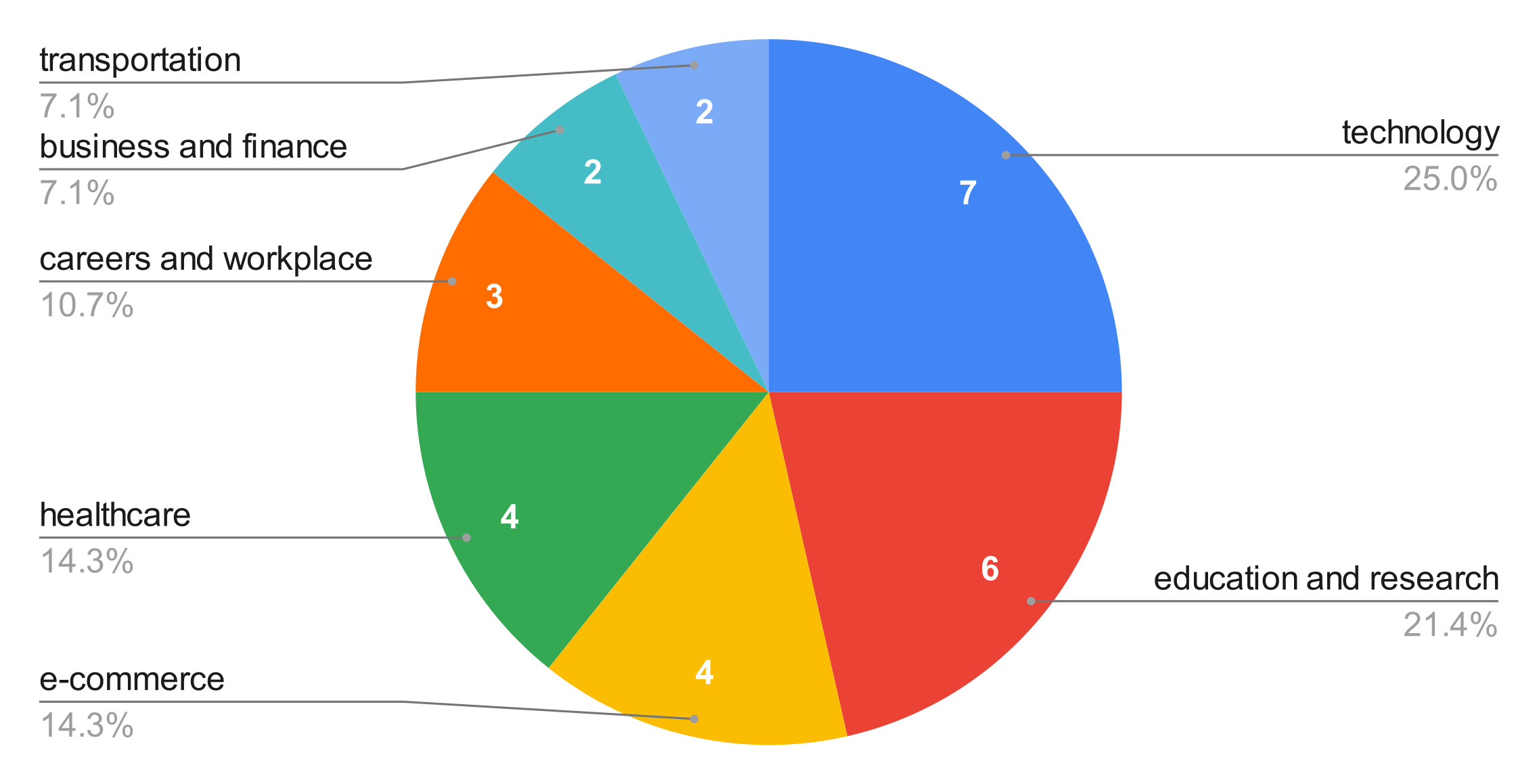}
    \caption{Domain distribution of card sorting studies in the analyzed dataset.}
    \label{fig:topics}
\end{figure}

The evaluation dataset comprises twenty-eight (28) open card sorting studies, with a total of 1,399 participants. The median number of cards in a study is 34 (\textit{IQR} = 24.75–49.25, \textit{MIN} = 10, \textit{MAX} = 87). The median number of participants in a study is also 34 (\textit{IQR} = 24.75–51, \textit{MIN} = 13, \textit{MAX} = 171). The weighted average number of categories created by participants across all studies, with weights adjusted to equalize the contributions of all studies independent of size, is 6.44 (\textit{SD} = 2.28). As shown in \autoref{fig:topics}, the analyzed studies offer thematic coverage of a variety of domains where card sorting is practiced, including technology (e.g., device and network configuration, security, accessibility), education and research, e-commerce, healthcare, career and workplace, business and finance, and transportation.

Geographically, participants were primarily located in the Western world at the time of completing the card sorting studies. Divided by continent, 839 were from North America (712 United States, 126 Canada), 448 from Europe (360 United Kingdom, 12 Netherlands), 80 from Oceania/Australia, 19 from Asia, 8 from Africa and 5 from South America. Unnamed countries were represented by 10 participants or fewer. Participants were recruited into the original card sorting studies in accordance with the validity criteria of the individual card sorting studies. Individual samples reflect the target audiences whose mental models were subject to investigation, and therefore contribute the ground truth for the purposes of our study. For example, an American e-commerce card sort concerning a category of products in a higher price range recruited homeowners from the USA in the high income range, with interest and hobbies aligned with said category of products, with equal gender representation and within the age range above 25 and below 71 years old. Another study targeted experts in a narrow professional field, without specific age or gender requirements (aside from being adults).

For analysis of the effects of study parameters (RQ4), studies were organized into terciles based on the number of cards, being Low ($\leq 26$), Medium (between 27 and 45) and High ($\geq 46$). The complexity of card labels was determined through collaborative coding (see \ref{sec:design} \hyperref[sec:design]{Experiment design and procedure}), resulting in a division of studies into three groups—Simple ($n=10$), Moderate ($n=9$), and Complex ($n=9$). The rating criteria included the conceptual difficulty and the level of domain-specific expertise required to understand the internal terminology in labels.

Card sorting studies include lists of cards, and their categorizations submitted by participants at the time when the studies were conducted. Their context further includes welcome messages and instructions used to communicate preliminary information to participants. Welcome messages were customized—elaborating goals, the domain, or other context—in twenty-two of the studies (79\%). Custom instructions were less common, applied in 9 of studies (32\%). They primarily expanded upon the physical procedure of sorting cards rather than its logical aspects. However, some instructions still discussed relevant context (e.g., asking participants not to use  certain logic while sorting the cards). In eight (8) studies, recruitment was conducted with UXtweak’s panel service. Therefore, specific targeting attributes (e.g., country, gender, household income) identify the sample demographics. Additionally, a single (1) among these studies included a custom screening question. The remaining twenty (20) studies did not explicitly specify participant eligibility criteria.

The analyzed card sorting studies contain protected information proprietary to select UXtweak users. The permission granted by the data owners through the informed consent allows us to process the data exclusively for the purpose of this research and the academic publication of its aggregated results in its anonymized form. Therefore, the original data—including card labels, categorization results and other artifacts—is not made public nor discussed in a manner that would enable the identification of participating data owners or the contents of their card sorting research.

\subsection{Card sorting simulation and data preprocessing}
\label{sec:simulation}

Data from card sorting studies (e.g., cards, context, number of participants) was used to construct LLM prompts shown in \ref{sec:prototype} \hyperref[sec:prototype]{Prototype technical implementation}. Sample demographics were established based on panel targeting attributes and qualifying answers in custom screening questions. All prompts were provided as inputs to LLM models separately to prevent carryover bias and to generate outputs in mutual isolation. 

With GPT-4o, results were obtained for all prompts (P1-P4), given the model’s capacity to handle large-volume CSV file outputs, which is crucial for prompts P1 and P2. Gemini, Claude, and Deepseek were determined through preliminary empirical evaluation as incapable of generating raw card sorting results or similarity matrices that would support further meaningful analysis. Therefore, only prompts P3 and P4 were run through these models. In total, including all studies, prompts, LLM models and repeated trials aimed at assessing the variability of outputs (4 times per GPT-4o prompt), the number of prompts executed and included in the final evaluation was 616.

Before evaluation, preprocessing was necessary due to errors that continued to manifest in generated outputs in spite of prompting strategies aimed at maintaining consistency and qualitative compliance. For example, a common error manifested as missing cards in the output. Initially, attempts were made to mitigate these issues through the development of an iterative chain-of-thought process where the LLM was prompted to address issues until the output was acceptable for standardized fully automated processing. However, new issues emerged as a result (e.g., the missing card issue was resolved, but the card was placed alone in an otherwise empty new category). Re-execution of the initial tuned prompt until its output was verified without errors was found to be the more efficient method, given that only a single prompt was typically needed, whereas chain-of-thought is normally resource-intensive. Generation of new outputs was necessary in 17.14\% of prompts, and was typically only needed once.

Apart from unifying the output format for real and synthetic data, the integrity of synthetic data was also a subject to verification. The following verification rules were applied to outputs of all prompts:

\begin{itemize}
    \item Every card must be present and sorted into a category.
    \item No card can be sorted into more than a single category.
    \item No new cards are introduced by hallucination.
    \item Two or more non-empty categories must be created.
\end{itemize}

Additionally, for P2, the similarity matrix must be symmetrical. Some outputs also required a manual correction, such as misformatted CSV files or inconsistent card names due to missing, replaced or added characters.

\subsection{Measures and evaluation process}

\begin{figure}[!ht]
    \centering
    \includegraphics[width=0.9\linewidth]{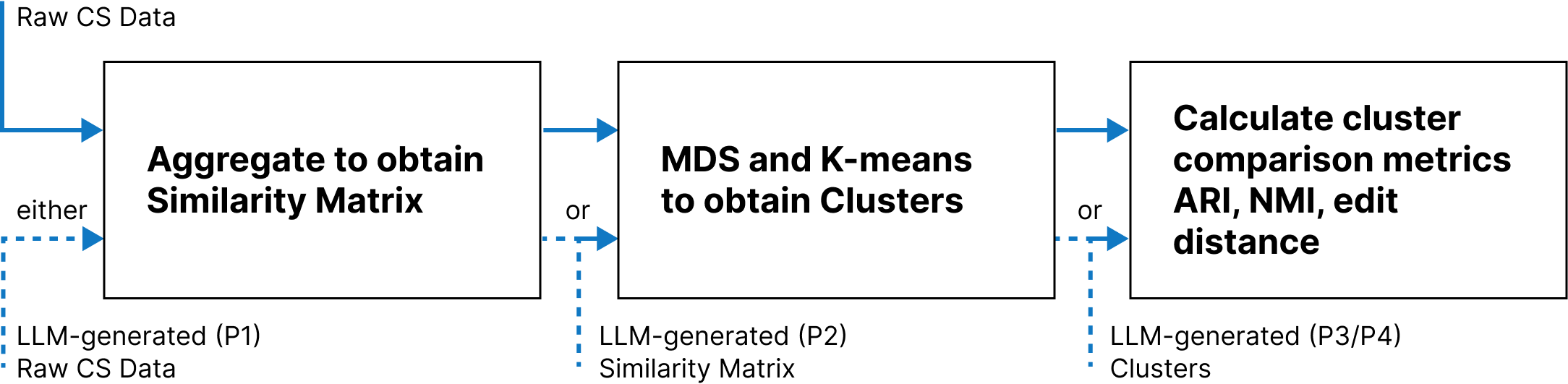}
    \caption{Simulated card sorting (CS) evaluation pipeline. LLM-generated and real card sorting results are processed identically, with simulated result data entering the pipeline at a step that reflects their initial state (raw data, similarity matrix, clusters), as determined by the LLM prompt’s output format.}
    \label{fig:eval-pipeline}
\end{figure}

The comparison of natural and simulated card sorting results (including all three experimental output format variants) required the application of consistent measures across all conditions. The output formats of the Card Sorting Simulator represent the results of open card sorting at three different levels of abstraction. Raw data (P1, as well as results from real participants) can be aggregated to form a similarity matrix (P2), from which clusters can be extracted (P3). Although result analysis can be approached from multiple angles, card clustering represents a key model for understanding the placement of each card within a grouping of related items \citep{martin2023}. Therefore, card sorting results were processed according to the evaluation pipeline to obtain comparable card clusterings as shown in \autoref{fig:eval-pipeline}. At the end of the pipeline, measures of similarity were calculated for clusterings to compare LLM-generated results to the ground truth results from real participants.

The similarity matrix (see \ref{sec:cardsort} \hyperref[sec:cardsort]{Card sorting user research method}) was constructed as a quantification of frequencies with which pairs of cards were grouped together by participants. Multidimensional scaling (MDS) was employed to map items into multidimensional space where similar items are close together, and dissimilar items are further apart. Subsequently, clustering was obtained through the K-means algorithm applied to card card coordinates in this low-dimensional space to identify commonly agreed-upon card groupings.

To find the optimal number of clusters, K was determined automatically by identifying the knee/elbow point of function $f$, which plots the number of clusters $K$ against the within-cluster sum of squares (WCSS). The WCSS is the sum of squared distances between data points and the centroids of their respective clusters. The knee/elbow point ($\kappa$) is defined as the point of maximum curvature on the curve of function $f$, calculated from the second and first derivative of the function as shown in \autoref{eq:curve}. Due to the discrete nature of function $f$, a continuous curve that approximately preserves the function’s original shape was obtained by applying smoothing splines and normalization of the function \citep{satopaa2011}.

\begin{equation}
    \label{eq:curve}
    \kappa_f(x) = \frac{f''(x)}{(1+(f'(x))^2)^{1.5}}
\end{equation}

Clustering similarity metrics applied in the evaluation include the number of categories (clusters), the Adjusted Rand Index (ARI), the Normalized Mutual Information Score (NMI) and edit distance \citep{zheng2020, cady2024}. The ARI is a robust measure for comparing clustering outcomes against random results, yielding values between -1 (indicating worse-than-random clustering) and 1 (perfect agreement). The NMI quantifies the amount of shared information between the clusterings (from LLM-generated and real participant results), making it a suitable primary measure for our evaluation, since it can robustly compare clusterings where the number of clusters differs. It returns a value between 0 (no mutual information) and 1 (perfect agreement). The {\ttfamily sklearn} Python library\footnote{Scikit-learn Python library: \url{https://scikit-learn.org/}} was used to calculate both the NMI and ARI measures. Edit distance is calculated as the minimum number of modifications required to transform one card sort into another \citep{nawaz2012}. Additionally, the number of categories is also compared.

Mantel tests were performed to assess correlations between similarity matrices, determining whether significant relationships exist between coupled sets of pairwise distances (or dissimilarities) between cards. The test was applied to the complement matrix, obtained by subtracting each matrix value from 100\% to express distances rather than similarities. For P1 and P2, similarity matrices were obtained naturally through the evaluation pipeline. Given an interest in thorough analysis, matrices were also created in reverse for P3 and P4. Although this involves mapping pairs of cards from the clustering (a single final categorization), so the matrices for P3 and P4 are inherently sparse, this approach enables the visual comparative analysis of patterns.

To account for the non-normal distribution of the data while comparing the values of metrics across the complete dataset of card sorting studies, non-parametric statistical tests were used. Kruskal-Wallis tests with Dunn’s post-hoc tests (with Holm’s correction) were applied for multiple group comparisons. The Wilcoxon signed rank test and Friedman test were used to assess differences across repeated measures or related samples (e.g., the same card sort evaluated with different models). The Friedman test was followed by the Nemenyi post-hoc test to pinpoint specific group differences.

A quantitative perspective does not fully capture the logical aspects of mental models investigated by card sorting. To obtain a fuller picture of the similarities and differences between natural human card sorting and the AI-driven approximation, we also conduct qualitative reviews.

\section{Results}
\label{sec:5}

Our combined quantitative and qualitative analysis investigates LLM-driven card sorting results to assess their properties under experimental conditions.

\subsection{General assessment of LLM-generated card sorting (RQ1)}

\textit{RQ1: Do LLM-generated card sorting results align with those produced by real participants?}

Even from the bird’s eye view, the number of clusters obtained through K-means highlights differences and similarities between the assessed methods. Card sorting studies with real participants—the ground truth method—yielded a median of 8 clusters (\textit{IQR} =  7–11). In P1, the number of clusters obtained by aggregating LLM-simulated raw card sorting results is considerably higher, with a median of 10 (\textit{IQR} = 8.75–12.25), which is significantly different according to the Willcoxon signed rank test, $W(n=54) = 36, p = .005, r=.54$. Over-segmentation indicates data that is more noisy or scattered. Similarity Matrix Generation with P2 yields a more similar number of clusters with a median of 8.5 (\textit{IQR} = 6–10.25, $W(n=54) = 142.5, p = .88$). Clustering Generation with P3 yields the most similar number of clusters to data from humans, with a median of 8 (\textit{IQR} = 5.75–9, $W(n=224) = 2075, p = .12$). 

These observations are further corroborated by comparing the agreement of clusterings via cluster- and matrix-based metrics, as shown in \autoref{tab:main-results}. The Friedman test of the NMI ($\chi^2(2, n=84) = 21.71, p < .001, W=.39$) supports statistically significant differences with medium effect, which post-hoc tests attribute to the difference of P1 from the other formats. P2 and P3 demonstrate similar degrees of moderate agreement that outperform P1. Given the values of agreement scores, structural similarities can be expected, but with some inconsistencies. On average, 11 to 12 cards would need to be modify their placements for an LLM simulation to obtain clusterings identical to a standard card sorting.

\begin{table}[!ht]
\centering
\caption{Overview of agreement measures comparing ground truth results of card sorting with real participants, and LLM (GPT-4o) simulations of results obtained through output format variants P1, P2 and P3. Bold text indicates the highest similarity.}
\begin{tabular}{lllllllll}
\toprule
\textbf{LLM output format} & \multicolumn{2}{l}{\textbf{NMI}} & \multicolumn{2}{l}{\textbf{ARI}} & \multicolumn{2}{l}{\textbf{Edit distance}} & \multicolumn{2}{l}{\textbf{Matrix correlation}} \\
\midrule
 & M & SD & M & SD & M & SD & M & SD \\
\midrule
\textit{P1} & 0.50 & 0.15 & 0.10 & 0.22 & 19.04 & 12.61 & 0.10 & 0.25 \\
\textit{P2} & \textbf{0.68} & 0.11 & 0.39 & 0.22 & 11.96 & 8.14 & \textbf{0.47} & 0.17 \\
\textit{P3} & \textbf{0.68} & 0.12 & \textbf{0.42} & 0.19 & \textbf{11.32} & 7.26 & 0.43 & 0.19 \\
\bottomrule
\end{tabular}
\label{tab:main-results}
\end{table}

\begin{figure}[!ht]
    \centering
    \includegraphics[width=0.8\linewidth]{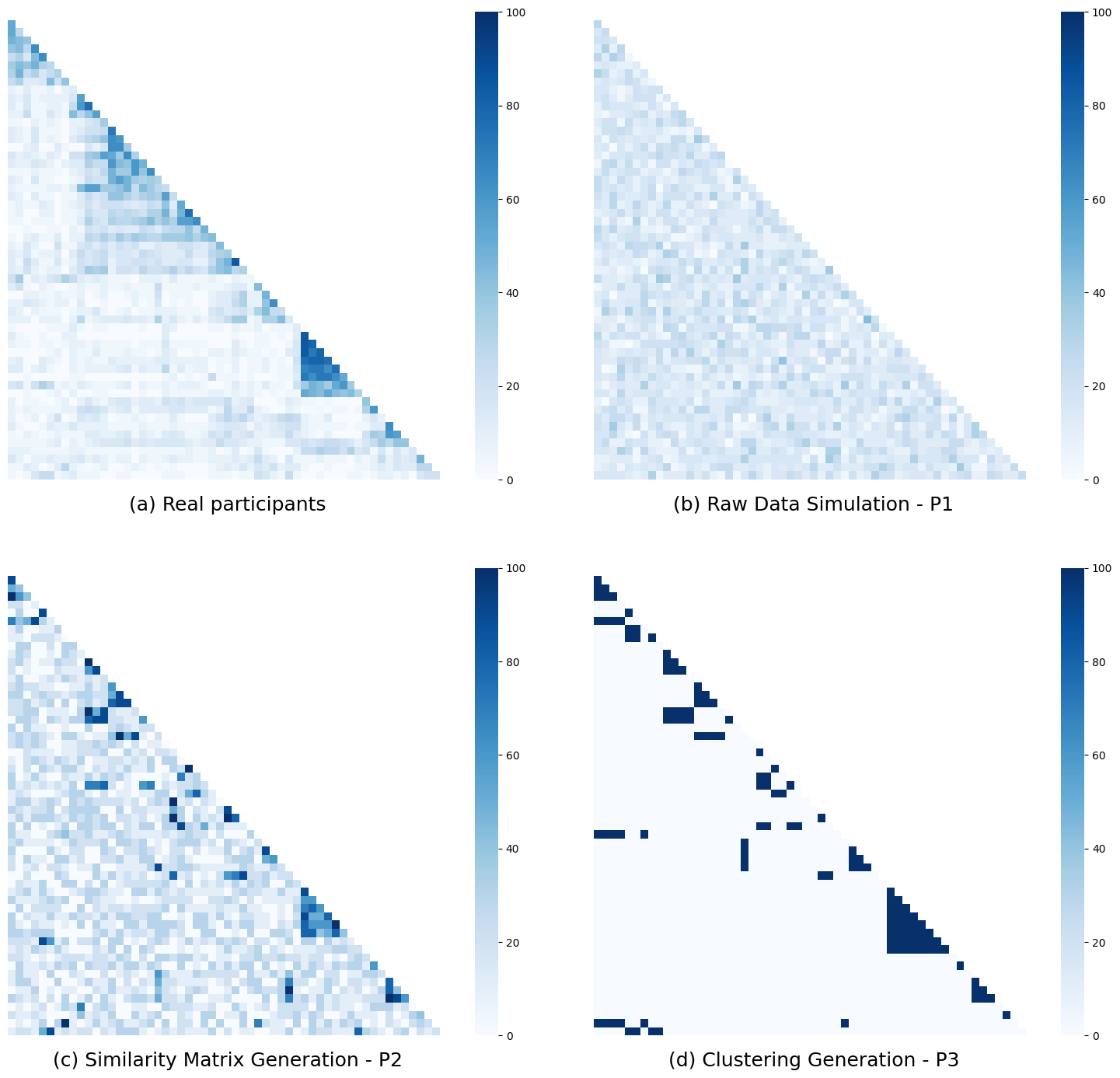}
    \caption{Example of similarity matrices calculated from real card sorting data and experimental generative methods of the Card Sorting Simulator, demonstrating structural divergence. GPT-4o was used for result generation.}
    \label{fig:similarity}
\end{figure}

To better understand the mutual relationship between the results of LLM-generated card sorting and data from real participants, we performed an in-depth analysis to compare the structure, conceptual features and labeling of the mental models being represented. Structural characteristics can be observed by comparing similarity matrices as seen in \autoref{fig:similarity}. Cards that real participants logically link together appear as darker shades on the diagonal. P1 either produced results that were too noisy, or in 5 out of 28 studies, completely identical for each simulated participant. P2 results are an improvement, since dark-shaded clusters of closely-related cards start appearing on the diagonal. However, medium shades of blue that indicate less common but still recurring patterns in real data are instead filled with random noise. P3 can yield groupings that are majorly similar to real data, but it can also create phantom groupings or assignments of individual cards that have significantly lesser prominence in real card sorting results.

Simulated card sorting also had a tendency to group cards into categories at conceptual levels either too generic or too specific in comparison to real participants. As an example of the former, in a study with a finance theme, the LLM simulator created a category titled “Financial operations” that included a significant portion of cards. By contrast, human participants created multiple smaller categories indicative of their expectations for operations that are logically linked, based on their understanding of these operations, previous experiences and potentially use of similar systems. As another example, a category titled “Sales \& marketing” was generated as an accumulation of commercial materials, but actual users distributed them among thematically associated groups. Furthermore, as is typical for open card sorting, real sorting delivered a significant diversity of category titles submitted by participants, even while inspecting the placements of a single card. Synthetic results did not reflect this diversity and the more high-level titles that it generated were commonly not found among participant submissions.

By contrast, overtly specific categories were also sometimes encountered in LLM results. This predominantly happened in cases where the card list contained smaller groups that could be linked together by narrow logical criteria. The LLM would prioritize keeping these cards in a single category together, even as human participants grouped them with other cards that were also linked to the same perceived logical concepts, albeit with more abstract or use-driven thinking.

To raise the internal validity of presented findings, we ensured they are not the results of external factors or random variation during non-deterministic black-box generation of responses by the LLM. We investigated the variability of generated results over multiple independent rounds. Each GPT-4o prompt for every card sorting study was executed four times. Variability of results of each output format was assessed as the mean of within-study standard deviation of the evaluated metrics (NMI, ARI, edit distance, matrix correlation). As demonstrated in \autoref{tab:variability}, the variability was negligible enough to eliminate random variables as a factor on the output. P3 exhibited the highest consistency, with the highest standard deviation of NMI within a study being .08 and the lowest being .004.

\begin{table}[!ht]
\centering
\caption{Overview of variability of clustering metrics across prompts for the GPT-4o model. Bold text indicates lowest variability.}
\begin{tabular}{lllp{3cm}p{3.8cm}}
\toprule
\textbf{LLM output format} & \textbf{NMI mean SD} & \textbf{ARI mean SD} & \textbf{Edit distance mean SD} & \textbf{Matrix correlation mean SD} \\
\midrule
\textit{P1} & 0.10 & 0.15 & 3.22 & 0.17 \\
\textit{P2} & 0.05 & 0.08 & 1.96 & 0.07 \\
\textit{P3} & \textbf{0.03} & \textbf{0.05} & \textbf{1.10} & \textbf{0.05} \\
\bottomrule
\label{tab:variability}
\end{tabular}%
\end{table}

Hypothesis H1a is therefore supported—among assessed approaches to card sorting simulation, Clustering Generation (P3) produces results that are the most accurate reflections of card sorting by humans. However, Similarity Matrix Generation yields results that are reasonably comparable. Hypothesis H1b is also supported, given the partial agreement between the LLM’s outputs and results from real participants.

\subsection{LLM model selection (RQ2)}

\textit{RQ2: How does the selection of the LLM model affect the similarity of LLM-generated card sorting results to data from real participants?}

Evaluations of the different LLM models based on the measures of agreement between clusterings from actual and simulated card sorting are shown in \autoref{tab:model-results}. Clustering Generation (P3) was applied from this assessment onward, given its position as the best performing output format in RQ1, as well as being universally applicable across investigated LLM models (owing to its compact output size that makes it compatible with lower maximum token limits).

\begin{table}[!ht]
\centering
\caption{Overview of LLM models applied for Clustering Generation (P3), compared to real card sorting results. Bold text highlights measures for best performing models (Claude for NMI, ARI and matrix correlation, Gemini for edit distance).}
\begin{tabular}{lllllllll}
\toprule
\textbf{LLM output format} & \multicolumn{2}{l}{\textbf{NMI}} & \multicolumn{2}{l}{\textbf{ARI}} & \multicolumn{2}{l}{\textbf{Edit distance}} & \multicolumn{2}{l}{\textbf{Matrix correlation}} \\
\midrule
 & M & SD & M & SD & M & SD & M & SD \\
\midrule
claude & \textbf{0.73} & 0.11 & \textbf{0.48} & 0.21 & 10.71 & 7.56 & \textbf{0.50} & 0.21 \\
deepseek & 0.68 & 0.14 & 0.41 & 0.24 & 11.57 & 8.08 & 0.42 & 0.23 \\
gemini & 0.72 & 0.11 & 0.46 & 0.20 & \textbf{10.43} & 7.91 & 0.48 & 0.20 \\
gpt4o & 0.68 & 0.12 & 0.42 & 0.19 & 11.32 & 7.26 & 0.43 & 0.19 \\
\bottomrule
\end{tabular}%
\label{tab:model-results}
\end{table}

Although Claude performed better than the other models in three out of four metrics, the overall differences are minimal. The Friedman test revealed statistically significant differences in NMI scores ($\chi^2(3, n=112) = 10.6, p = .014, W=.13$) and similarity matrix correlation ($\chi^2(3, n=112) = 9.32, p = .025, W=.11$), both with small effects. Post-hoc tests revealed that the only significant differences were between Claude and DeepSeek ($p=.022$ for NMI and $p=.047$ for correlations). Therefore, hypothesis H2 is partially supported, considering only a singular instance of small statistically significant difference.

\subsection{Context inclusion (RQ3)}

\textit{RQ3: How does inclusion of context in the prompt affect the similarity of LLM-generated card sorting results to data from real participants?}

Comparison of Clustering Generation with and without context revealed no significant differences as shown in \autoref{tab:context}. Post-hoc tests to a Friedman test involving all four output format variants ($\chi^2(3, n=112) = 25.23, p < .001, W=.3$) indicate that P4 only differs from P1, in a manner identical to P3.

\begin{table}[!ht]
\centering
\caption{Overview of agreement between results of Clustering Generation with and without context (P3 and P4 respectively) and real card sorting results. Bold text indicates the highest similarity.}
\begin{tabular}{lllllllll}
\toprule
\textbf{LLM output format} & \multicolumn{2}{l}{\textbf{NMI}} & \multicolumn{2}{l}{\textbf{ARI}} & \multicolumn{2}{l}{\textbf{Edit distance}} & \multicolumn{2}{l}{\textbf{Matrix correlation}} \\
\midrule
 & M & SD & M & SD & M & SD & M & SD \\
\midrule
\textit{P3} & 0.68 & 0.12 & \textbf{0.42} & 0.19 & \textbf{11.32} & 7.26 & \textbf{0.43} & 0.19 \\
\textit{P4} & \textbf{0.69} & 0.11 & \textbf{0.42} & 0.19 & 11.93 & 7.54 & \textbf{0.43} & 0.18\\
\bottomrule
\end{tabular}%
\label{tab:context}
\end{table}

\begin{figure}[!ht]
    \centering
    \includegraphics[width=0.7\linewidth]{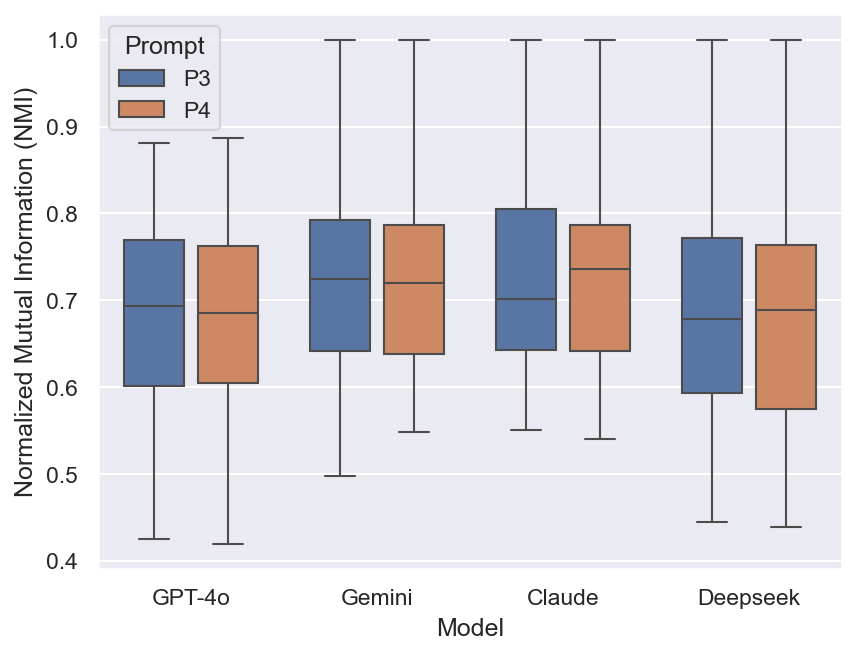}
    \caption{Comparison of Normalized Mutual Information scores across four LLMs. Visually, boxplots show only negligible differences between models.}
    \label{fig:models}
\end{figure}

Building upon the findings from RQ2 that there are only minimal differences in the results obtained with different LLM models while context is provided, we also investigated whether any of the previously assessed LLM models (Claude, DeepSeek, Gemini, GPT-4o) would be more or less sensitive to the removal of context. This comparison is visualized in \autoref{fig:models}. For the P4 prompt, Friedman tests did not reveal statistically significant differences between models for NMI ($\chi^2(3, n=112) = 6.95, p = .074$) or for matrix correlation ($\chi^2(3, n=112) = 7.67, p = .053$). This is only a small change against P3, where a small difference was found between DeepSeek and Claude, favoring the latter. Therefore, it can be suggested that context helps Claude—the best model during our investigation—achieve results that are more reflective of card sorting from real participants, but only slightly, since the difference is significant only against DeepSeek.

The demographic attributes that defined the card sorting sample were only available in eight (29\%) of the studies in the dataset. Since description of the sample (participant variables) could be viewed as critical for generating realistic card sorting results, we separately evaluated for the effects of context for this subset only. However, statistically significant differences were not found between P3 and P4 in any of the measures. NMI score: $W(n=16) = 12, p=.4$. Similarity matrix correlation: $W(n=16) = 9, p=.25$. To identify potential conditions where context may still contribute to more accurate simulated card sorting, we inspected studies for outliers that could serve as counterexamples, but no such studies could be identified.

Therefore, hypothesis H3 is disproved. Context obtained from card sorting studies as represented in our prompts (demographic attributes of the sample, text messages) does not inherently lead to simulated card sorting results that would be more similar to sorting by real participants.

\subsection{Effects of study complexity (RQ4)}

\textit{RQ4: How does the complexity of a card sorting study (number of cards, complexity of card labels) affect its LLM-generated results?}

\begin{figure}
    \centering
    \includegraphics[width=\linewidth]{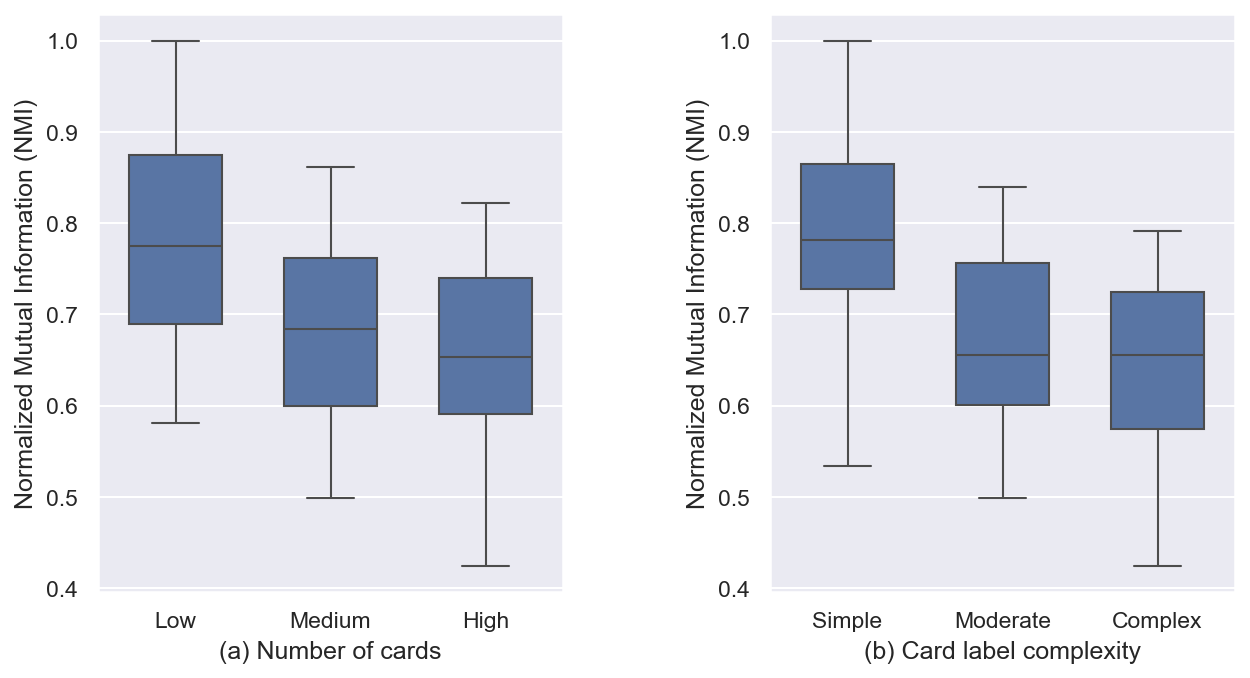}
    \caption{Comparison of Normalized Mutual Information scores across three card count levels (a) and three card label complexity levels (b). Higher numbers of cards increase the scale of deviations between the simulations and real card sorting results, while simple card labels result in simulations that are most similar to card sorting by human users.}
    \label{fig:cards-diff}
\end{figure}

The number of cards as an indicator of the complexity of a card sorting study has a significant link to the accuracy of simulated card sorting results. The mean NMI score for the Cluster Generation (P3) prompt is the highest for studies with a low card count (\textit{M}=0.77, \textit{SD} = 0.14), followed by medium card count (\textit{M}=0.7, \textit{SD} = 0.1), while high card count studies yielded results that were the most different from those obtained from real participants (\textit{M}=0.65, \textit{SD} = 0.09). This is confirmed by applying a Kruskal-Wallis test, $H(n=112) = 19.1, p < .001, \eta^2 = .16$. The negative relationship between the number of cards and the accuracy of simulated card sorting is demonstrated by NMI as seen in \autoref{fig:cards-diff}.

Card label complexity affects the degree in which results of the simulation agree with data from real participants. The average NMI score for simulations with the P3 prompt is \textit{M}=0.78 (\textit{SD} = 0.13) for studies with simple card labels, \textit{M}=0.68 (\textit{SD} = 0.09) for moderate, and \textit{M}=0.64 (\textit{SD} = 0.1) for difficult card labels. The Kruskal-Wallis test confirms these differences as statistically significant, $H(n=112) = 26.14, p < .001, \eta^2 =.22$. A visual comparison of study groupings by label complexity based on NMI is shown in \autoref{fig:cards-diff}.

\begin{figure}[!ht]
    \centering
    \includegraphics[width=0.9\linewidth]{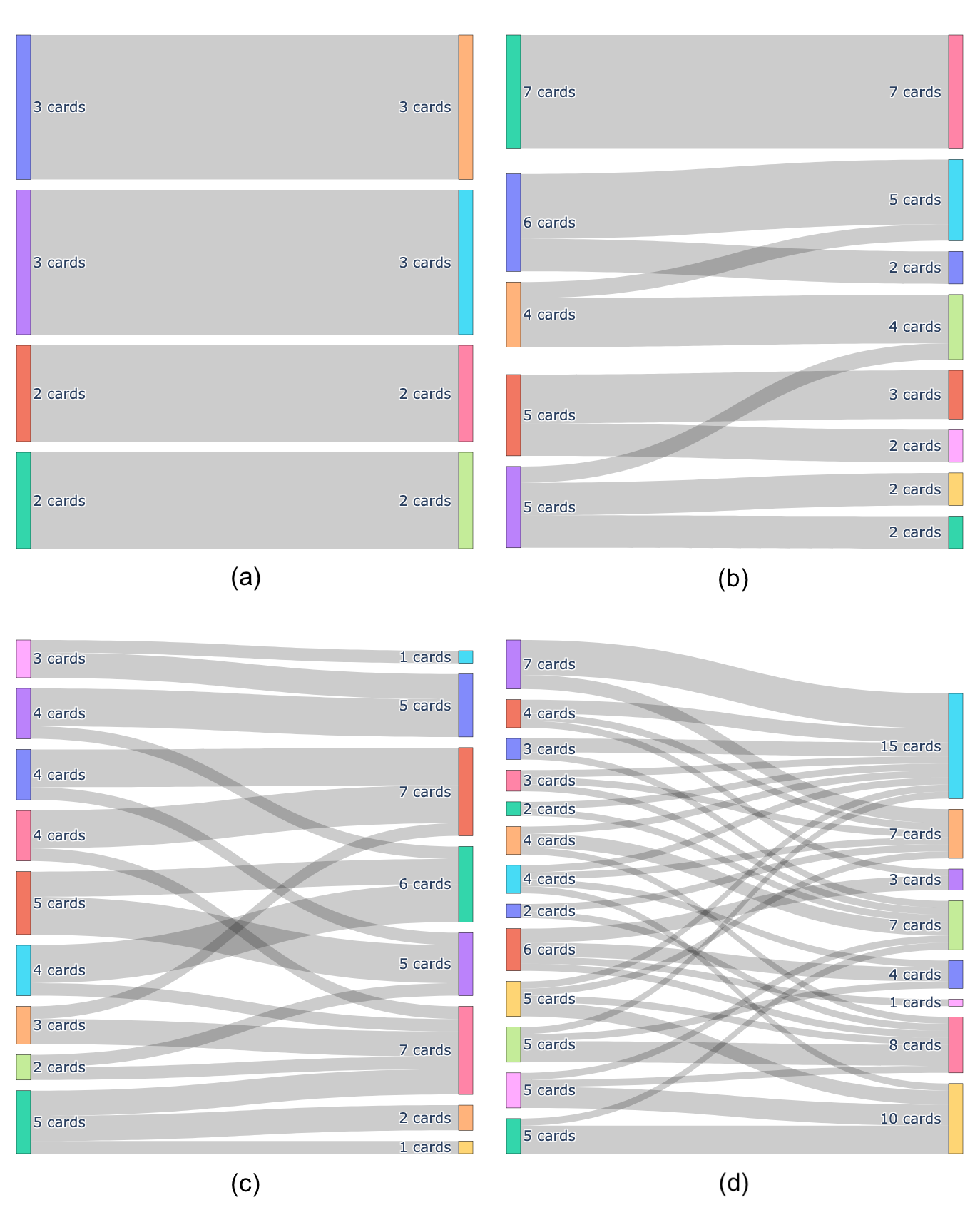}
    \caption{Comparison of card clusterings between examples of studies with different complexity (a) 10 cards, simple labels, NMI = 1 (b) 27 cards, simple labels, NMI = 0.8, (c) 34 cards, moderate labels, NMI = 0.6, (d) 55 card, complex labels, NMI = 0.42. Clusterings from real participant results are on the left, synthetic clusterings are on the right.}
    \label{fig:mapping}
\end{figure}

In-depth inspection corroborates agreement between real and synthetic results as a product of the two complexity indicators. Studies with the highest agreement NMI of 1 were generally simple studies with small numbers of cards, while the largest score of 0.42 was discovered in a study with a high number of cards (55) and complicated labels. Contrasts between the mappings is shown in \autoref{fig:mapping}.

Therefore, our findings support hypothesis H4. When a card sorting study is more complex, it is likely that its LLM simulation will result in deeper inconsistencies against mental models of real participants than a study that comprises simpler information.

\subsection{Output errors (RQ5)}
\label{sec:errors}

\textit{RQ5: What are the common issues that appear when LLMs are used to simulate card sorting?}

Several types of errors relevant for effective card sorting simulation occurred within the outputs of all LLM models, particularly in card sorting studies with high card counts. Most prevalent errors can be summarized as mismanagement of cards, such as omission of expected cards or presence of duplicates. A recurring pattern was that the model concluded its response with statements such as: "All X cards were included with no duplicates or omissions." in spite of them sometimes being incorrect. This is in spite of explicit directives to verify the number of cards. 

In some instances, because of the validation directives, the model corrected itself iteratively. This could repeat up to six times within a single prompt, before arriving at the expected output. However, this behavior was inconsistent in occurrence and effectiveness. Examples of such self-corrections include:
\begin{itemize}
    \item "It looks like there is a mismatch between category names. I’ll fix that and regenerate the similarity matrix."
    \item "It looks like I have one extra card in my categorization. I’ll review the categories and ensure exactly 57 cards are included without duplicates. Let me correct this."
    \item "It looks like some cards were missed in the categorization. I’ll review and correct the categorization to ensure all 42 cards are included exactly once."
\end{itemize}

Other errors involved modifications to card names, such as grammatical corrections, adjustments for consistency, or the replacement of punctuation. Additionally, formatting issues were observed, including the introduction of extra rows or unintended spaces in tabular outputs. Notably, no models hallucinated new cards into existence, even in P1 and P2 as the more complex prompts. See \autoref{tab:problems} for a complete list of identified output problems.

\begin{table}[!ht]
\caption{Overview of errors identified in LLM outputs. Items are sorted by significance for use and processing. Errors 8-1, 8-2 and 8-3 are subtypes of error 8 listed individually due to their frequency.}
\resizebox{\columnwidth}{!}{%
\begin{tabular}{llp{7cm}lll}
\toprule
\textbf{\#} & \textbf{Problem} & \textbf{Description} & \textbf{Affected models} & \textbf{Prompts} & \textbf{Frequency order} \\
\midrule
1 & Omission of expected cards & One or more cards are missing from the output, leading to an incomplete result. & All & All & 1 \\
2 & Redundant card inclusion & Duplicate copies of one or more cards are introduced, sometimes placed into multiple categories at once. & GPT-4o & P1, P3, P4 & 6 \\
3 & Overcategorization & Almost all cards are placed into separate categories by themselves, precluding the purpose of card sorting. & DeepSeek & P3, P4 & 9 \\
4 & Undercategorization & All cards are grouped into one category, precluding the purpose of card sorting. & DeepSeek & P3, P4 & 8 \\
5 & Asymmetrical matrix & The similarity matrix is not symmetrical as required. Exclusive to P2. & GPT-4o & P2 & 5 \\
6 & Looping & The LLM enters an infinite/long recalculation loop while attempting to resolve another error (Omission of expected cards). & GPT-4o, DeepSeek & All & 7 \\
7 & Output not printed & The task is declared completed, yet the result is not shown in the actual output. & GPT-4o & All & 4 \\
8 & Card label modification & Card labels are modified, which may alter their intended meaning. & GPT-4o, DeepSeek & All & 2 \\
8-1 & Space removal & Extra spaces in card labels are removed. & DeepSeek, Claude, GPT-4o & All & 2 \\
8-2 & Punctuation removal & Punctuation is removed, such as periods at the end of card labels. & DeepSeek & P3, P4 & 2 \\
8-3 & Special character replacement & Specific characters are replaced—e.g., typographic apostrophes (’) for their straight equivalents (')—impacting consistency. & Claude & P3, P4 & 2 \\
9 & Extra whitespace & Unnecessary line breaks and spaces create empty lines or white space in the CSV output. & Gemini & P3, P4 & 3 \\
\bottomrule
\end{tabular}%
}
\label{tab:problems}
\end{table}

To answer hypothesis H5, there are errors in LLM outputs that reduce their reliability for card sorting simulation, requiring cautious validation and the implementation of additional processing strategies such as multiple prompting and output post-processing. The most prevalent issues persisted among all LLMs throughout the experiment, although GPT-4o and DeepSeek demonstrated a greater (more disruptive) error diversity than Gemini and Claude.

\section{Discussion}
\label{sec:6}

The findings from our investigation support the concept of LLM card sorting and the proposed Card Sorting Simulator as a component of AI-augmented instrument for information architecture research. Significant differences from mental models of real users also serve as a warning sign against potential attempts at leveraging AI as a fully automated replacement for real human feedback. However, moderate agreement about the most prominent card clusters in the similarity matrix indicates potential usefulness for the processes involved in data collection and analysis as part of the proposed instrument (e.g., fast preliminary feedback, comparative analysis with obtained data).

\subsection{Implications}

The struggle of the LLMs to emulate real-world diversity is particularly evident in synthesis of raw results, which either resemble random noise without logical patterns, or produce monoliths, where each result lacks any realistic individual perspective. Generated similarity matrices effectively combine these aspects, containing some prominent clusters determined as most salient by the LLM, but with the majority of the matrix remaining random. Thus, more nuanced or less common conceptual models are neglected. Therefore, logical result diversity, as well as the capacity to capture more contextually specific and experience-driven mental models present key challenges for the improvement of card sorting result synthesis.

Among the evaluated models, there is no LLM that can be clearly declared as the best or the worst based on its accuracy in simulating the results of real-world card sorts within our dataset. Claude and Gemini can be considered only slightly more preferable over GPT-4o and DeepSeek, given that their results are less polluted with errors. Notably, since LLM models are black boxes that continue to evolve, some of these errors may be resolved in the future, or new ones may be introduced. Monitoring models will be justified in future development. High consistency between the different models highlights the cross-model generalizability of presented findings. 

For effective and efficient card sorting simulation, it is relevant that neither the more detailed output data formats, nor the addition of context aside from the list of cards had significant positive impacts on the accuracy of the synthetic results. Furthermore, Raw Data Simulation had a negative effect. While Similarity Matrix Generation has similar accuracy to Clustering Generation, it is significantly more complex, involving higher numbers of output tokens, which makes Clustering Generation the most optimal choice among the investigated approaches.

In this research, we focused on open card sorting (see \ref{sec:cardsort} \hyperref[sec:cardsort]{Card sorting user research method}) due to its higher complexity and greater potential for diverse results. In closed card sorting, participants (or LLM-driven tools synthesizing their results) have less freedom to represent mental models. In hybrid card sorting, they are guided to think in a certain way. The three card sorting variants have different methodological purposes and are used in different stages of information architecture design. Therefore, direct comparisons of simulation accuracy between the three approaches would not be of substantial relevance. While LLMs might perform better in closed or hybrid card sorting with pre-existing categories as a template, reducing the impact of self-expression, it may still be vulnerable to similar issues such as lack of diversity. New issues may also arise from users and LLMs interpreting pre-existing categories differently, or divergence in needs to create new categories.

The ability of the Card Sorting Simulator to accurately reflect mental models could be be improved by utilizing more task-specific machine learning models instead of LLMs trained purely on general datasets. By training custom LLM models on data from real card sorting studies, or utilizing Reinforcement Learning from Human Feedback \citep{stiennon2020}, the models could be tuned to address problems such as simulation of diverse mental models and realistic levels of abstraction. However, given the variability of the types of information (card sets) and the potential participant samples, this could introduce further biases caused by different contexts of card sorting studies involved in such tuning.

Additionally, in the hypothetical event of high model convergence and saturated training data, there would still be a need for validation with real users due to a variety of factors, such as real-world diversity, inconsistency between the expected and actual characteristics of the intended audience, introductions of novel concepts, or dynamic changes in user perceptions over time. As a deep learning model, LLMs generate their output by drawing on their training data (typically, text from the Internet). By contrast, participants can draw on their experience and apply higher cognitive functions to form mental models in a relevant context. Simulated card sorting could be more analogous to card sorting by experts due to its potential bias, albeit leveraging links between tokens learned in a different context as opposed to actual expertise. By this logic, AI-driven card sorting simulation is within the same methodological paradigm as usability transpection, a term introduced by \citet{kuric2025slr} for usability assessment methods that depend on reapplying information learned automatically in different contexts rather than actual experience, expertise or heuristics.

\subsection{Limitations}

The context utilized in our experiment to provide LLMs with further details about card sorting studies comprised information that was readily available (messages, participant targeting attributes). This approach was used to assess the effectiveness of the instrument without introducing additional requirements on the user conducting the card sort. It is possible that by manually crafting context for the LLM, the quality of synthetic results could be improved. For example, it would be possible to manually craft personas for individual participants and aggregate their simulations from multiple prompts. However, this would require additional labor on the part of the card sorting researchers to expertly set up, make the method more computationally complex and costly, and potentially amplify the researcher’s biases. The usability of such a complex setup would hinder its use as rapid augmented feedback. Recruiting actual human participants who provide genuine human perspectives could be more efficient at that point. Nonetheless, for better understanding of the impact and value of context on the simulation of card sorting as a non-trivial cognitive task, future studies should investigate methods enabled by deeper contextual representations.

The selection of studies contained within the dataset could limit the external validity of our findings. To address such validity threats, the dataset contains card sorting studies of various complexities, conducted by UX researchers from various domains and covering varied topics. All studies were conducted in UXtweak as a standard online UX research tool. Nonetheless, it is possible that in a different sample of card sorting studies, synthetic results could manifest distinct phenomena. The card sorts were conducted—and the participants were recruited—primarily in Western countries (North America, Europe and Australia), which can be attributed to our focus on studies in the English language and performed in a card sorting tool with a user base established primarily in these countries. This could have introduced cultural, socio-economical and other biases into our study. To lend further generalizability to the empirical understanding of LLM-driven card sorting simulation, replication with different datasets will be beneficial.

Only a single study contained a custom screening question. Screening questions can be a source of information on participant inclusion/exclusion criteria that also offer inherent value, making them a source of context about participants that is less intrusive than researchers having to provide added context exclusively for the purposes of the LLM simulation module.  However, screening questions in card sorting studies can be non-existent or minimalistic, as evident from our dataset of card sorting studies conducted in practice. As a solution, experimental card sorting instruments could include verbalization of screening criteria as a non-optional step to encourage more methodologically rigorous recruitment. In this context, more detailed custom screening criteria could also be assessed.

As part of experimental evaluation, LLM outputs needed to be partially manually pre-processed due to inconsistent format and errors analyzed in section \ref{sec:errors} \hyperref[sec:errors]{Output errors}. While manual adjustments are manageable under experimental conditions, and they do not contribute any differences to card sorting semantics, for practical applications, robust normalization methods will need to be implemented.

\subsection{Future work}

The scope of this research spans the proposal of a system for AI-augmented card sorting and the empirical evaluation of the concept of the Card Sorting Simulator, which is its central AI-driven component. Given the positive findings, future steps for the development of the system will involve the integration of the Card Sorting Simulator into an online card sorting tool, design of user interfaces for generating, processing and analyzing its output, as well as the evaluation of its impact on card sorting research processes. Key signals of success will include the satisfaction of UX researchers with the tool, improvements to work efficiency and quality of the card sorting results and information architectures created as the outcome.

The theoretical understanding of LLM-driven card sorting simulation could be further enhanced by assessing simulations in the context of different types of card sorting, such as closed or hybrid card sorting, or with card sorting involving images either on their own, or supported by text.

Aside from the proposed Card Sorting Simulator, future research could also investigate other AI-driven augmentation techniques for information architecture research and design tools. For example, automated construction of information hierarchies based on the outputs of card sorting studies could reduce the gap between ideation and validation.

\section{Conclusion}
\label{sec:7}

This article proposed a system that incorporates AI augmentation into card sorting to aid UX researchers and other professionals with the planning and evaluation of studies aimed at the design of intuitive information architectures. To evaluate our approach, we implemented a prototype of an LLM-driven Card Sorting Simulator and performed its rigorous comparative analysis against the standard card sorting method that involves real human participants. A dataset of 28 different card sorting studies, previously conducted in real practice by independent researchers across diverse companies, organizations and industries was analyzed to produce findings of exceptional ecological validity. 

Our findings demonstrate that LLM models (e.g., Gemini, Claude, GPT-4o, DeepSeek) can simulate card sorting results accurately enough to provide quick preliminary feedback to researchers. At the same time, divergences from real mental models mount with increased study complexity, and the models have limited ability to account for real user diversity, discouraging the development of fully automated card sorting. We posit that this validates the concept of our proposed augmented system. Our research provides further motivation for exploring AI in the role of an assistant to alleviate some of the challenges encountered in the process of ideation of evaluation of information architectures.

\appendix
\section{Prompt templates}
\label{app:templates}

\subsection{Raw Data Simulation Prompt (P1)}

{\footnotesize \sffamily \setlength{\parindent}{0pt}
Instructions:\\
- Imagine you represent {\ttfamily\{numberOfParticipants\}} participants in an open card sorting study.\\
- Perform a card sort as these participants would, by grouping provided cards into meaningful categories based on their similarities.\\
- Assign descriptive names to categories so that they reflect the shared themes or characteristics of the cards within them. \\
- Output a list of all categories with their corresponding cards contained within for each participant.\\
- Treat each participant as a separate person, with their own individual, but still meaningful results based on the participant’s perspective and similarities between cards (e.g. different users may, in some cases, assign the same card under different categories).\\
- Use a reasonable number of categories for each user (more than one).\\
- For each participant, use all of the provided cards exactly once. \\
- Keep the cards as they are, including any grammatical mistakes or unnecessary characters.\\
- If text in brackets appears in the card list, it presents a description of the card that is provided for context only. This description should not appear in the results.\\

Group the cards from the perspective of participants with the following context:\\
- Demographic attributes for respondents are: {\ttfamily\{demographicAttributes\}}\\
- Welcome message received by participants is: "{\ttfamily\{welcomeMessage\}}"\\
- Instructions before the task received by participants are: "{\ttfamily\{instructions\}}"\\
- Additional considerations: all categories have to be named (no 'Unnamed category' entries)\\
\noindent
Output rules:\\
- Present your results in CSV format, with three columns: "respondent" (participant's id), "category" (name of the category) and "card" (card name).\\
- Enclose card and category names in quotation marks. If a quotation mark appears within the text, escape it by doubling it (""), as standardly done in CSV.\\
- Do not use unnecessary blank lines to divide categories.\\

Example:\\
In the example below, respondent 1 sorted Bird into Animals category while respondent 2 sorted Bird into Flying objects category:\\
{\`{}}{\`{}}{\`{}}  \\
{\ttfamily respondent,category,card\\
1,Animals,Bird\\
1,Vehicles,Plane\\
2,Flying objects,Bird\\
2,Flying objects,Plane}\\
{\`{}}{\`{}}{\`{}}\\

Cards to sort:\\
{\`{}}{\`{}}{\`{}}\\
{\ttfamily\{cards\}}\\
{\`{}}{\`{}}{\`{}}\\

Before displaying the CSV result, check if all cards are included exactly once - {\ttfamily\{numberOfCards\}} cards should be present for each participant.
}

\subsection{Similarity Matrix Generation Prompt (P2)}

{\footnotesize \sffamily \setlength{\parindent}{0pt}
Instructions:\\
- Imagine you represent {\ttfamily\{numberOfParticipants\}} participants in an open card sorting study.\\
- Generate a similarity matrix from provided cards, where rows and columns represent cards, and each cell indicates the percentage of participants who grouped the specific pair of cards together, based on the card similarities and shared characteristics, as well as the participants’ varied perspectives and logic.\\
- Keep the similarity values realistic, in the range from 0 to 100.\\
- To reflect the percentage of participants who paired them, assign high values to card pairs that are often grouped together and share similar themes. Assign low values to logically distant pairs that few participants would pair together.\\
- Values should represent fractions of the total number of participants (e.g., with four participants, valid values are 0, 25, 50, 75 and 100).\\
- Use all of the provided cards exactly once. \\
- Keep the cards as they are, including any grammatical mistakes or unnecessary characters.\\
- If text in brackets appears in the card list, it presents a description of the card that is provided for context only. This description should not appear in the results.
\\

Group the cards from the perspective of participants with the following context:\\
- Demographic attributes for respondents are: {\ttfamily\{demographicAttributes\}}\\
- Welcome message received by participants is: "{\ttfamily\{welcomeMessage\}}"\\
- Instructions before the task received by participants are: "{\ttfamily\{instructions\}}"\\
- Additional considerations: all categories have to be named (no 'Unnamed category' entries)\\

Output rules:\\
- Present your results in CSV format, where rows and columns in the matrix represent distinct cards.\\
- The matrix must be symmetrical and have zeros on the diagonal.\\
- Enclose card and category names in quotation marks. If a quotation mark appears within the text, escape it by doubling it (""), as standardly done in CSV. \\

Example:\\
In the example below, 40\% of participants placed Bird and Plane in the same category:\\
{\`{}}{\`{}}{\`{}}\\
{\ttfamily ,Cat,Dog,Bird,Plane\\
Cat,0,95,65,0\\
Dog,95,0,75,15\\
Bird,65,75,0,40\\
Plane,0,15,40,0}\\
{\`{}}{\`{}}{\`{}}\\

Cards to sort:\\
{\`{}}{\`{}}{\`{}}\\
{\ttfamily\{cards\}}\\
{\`{}}{\`{}}{\`{}}\\

Before displaying the CSV result, check if all cards are included exactly once - {\ttfamily\{numberOfCards\}} cards should be present, and the matrix must be symmetrical.
}

\subsection{Clustering Generation Prompt (P3)}

{\footnotesize \sffamily \setlength{\parindent}{0pt}
Instructions:\\
- Imagine you represent {\ttfamily\{numberOfParticipants\}} participants in an open card sorting study.\\
- Group the cards provided to you into meaningful categories based on their similarities.\\
- Assign descriptive category names that reflect the shared themes or characteristics of each category’s cards, while incorporating the perspectives and logic of the participants.\\
- Output a list of all categories with their corresponding cards. Your output is an aggregation, obtained after the individual results from all participants were analyzed and a single “best” card sorting result was created that aligns with collective mental models of users as best as possible.\\
- Use a reasonable number of categories (more than one).\\
- Use all of the provided cards exactly once, each card has to belong to one category only. \\
- Keep the cards as they are, including any grammatical mistakes or unnecessary characters.\\
- If text in brackets appears in the card list, it presents a description of the card that is provided for context only. This description should not appear in the results.\\

Group the cards from the perspective of participants with the following context:\\
- Demographic attributes for respondents are: {\ttfamily\{demographicAttributes\}}\\
- Welcome message received by participants is: "{\ttfamily\{welcomeMessage\}}"\\
- Instructions before the task received by participants are: "{\ttfamily\{instructions\}}"\\
- Additional considerations: all categories have to be named (no 'Unnamed category' entries)\\

Output rules:\\
- Present your results in CSV format with two columns: "categoryName" and "cardName".\\
- Enclose card and category names in quotation marks. If a quotation mark appears within the text, escape it by doubling it (""), as standardly done in CSV.\\
- Do not use unnecessary blank lines to divide categories.\\

Example:\\
In the example below, Cat and Dog were placed in the same category (Animals):\\
{\`{}}{\`{}}{\`{}}\\
{\ttfamily categoryName,cardName\\
Animals,Cat\\
Animals,Dog\\
Means of transport,Plane}\\
{\`{}}{\`{}}{\`{}}\\

Cards to sort:\\
{\`{}}{\`{}}{\`{}}\\
{\ttfamily\{cards\}}\\
{\`{}}{\`{}}{\`{}}\\

Before displaying the CSV result, check if all cards are included exactly once - {\ttfamily\{numberOfCards\}} cards should be present. No duplicates are allowed.
}

\subsection*{Declaration of competing interests}
The authors declare that they have no known competing financial interests or personal relationships that could have appeared to influence the work reported in this paper.

\subsection*{Funding sources}
This work was supported by the Slovak Research and Development Agency under Contract No. APVV-23-0408 and co-financed by the Cultural and Educational Grant Agency of Slovak Republic (KEGA) under Grant No. KG 014STU-4/2024. We would like to thank UXtweak j.s.a. for their generous financial support of this research and for the technical and expert support provided by the UXtweak Research team.

\subsection*{Data statement}
The authors do not have permission to publicly share the data. However, certain anonymized data may be made available on request.

\bibliography{sources}

\begin{thebibliography}{73}
\providecommand{\natexlab}[1]{#1}
\providecommand{\url}[1]{\texttt{#1}}
\expandafter\ifx\csname urlstyle\endcsname\relax
  \providecommand{\doi}[1]{doi: #1}\else
  \providecommand{\doi}{doi: \begingroup \urlstyle{rm}\Url}\fi

\bibitem[Tankala and Sherwin(2024)]{tankala2024}
Samhita Tankala and Katie Sherwin.
\newblock Card sorting: Uncover users' mental models for better information architecture.
\newblock \url{https://www.nngroup.com/articles/card-sorting-definition/}, 2024.

\bibitem[Baxter et~al.(2015)Baxter, Courage, and Caine]{baxter2015}
Kathy Baxter, Catherine Courage, and Kelly Caine.
\newblock \emph{Understanding Your Users: A Practical Guide to User Research Methods}.
\newblock Morgan Kaufmann, 2015.
\newblock URL \url{https://doi.org/10.1016/C2013-0-13611-2}.

\bibitem[Lewis and Sauro(2021)]{lewis2021}
James~R. Lewis and Jeff Sauro.
\newblock \emph{Usability and user experience: Design and evaluation}, chapter~38, pages 972--1015.
\newblock John Wiley \& Sons, Ltd, 2021.
\newblock ISBN 9781119636113.
\newblock URL \url{https://doi.org/10.1002/9781119636113.ch38}.

\bibitem[Munim et~al.(2024)Munim, Islam, Milton, Ara, Faisal, and Islam]{munim2023}
Kazi~Md. Munim, Iyolita Islam, Md. Musfiqur~Rahman Milton, Laila~Arzuman Ara, Faiz~Al Faisal, and Muhammad~Nazrul Islam.
\newblock Exploring the impact of design technique on usability: A case study on designing the ehealth websites using card sorting and interactive dialogue model.
\newblock \emph{Engineering Reports}, 6\penalty0 (3):\penalty0 e12738, 2024.
\newblock URL \url{https://doi.org/10.1002/eng2.12738}.

\bibitem[Paea et~al.(2022)Paea, Katsanos, and Bulivou]{paea2022}
Sione Paea, Christos Katsanos, and Gabiriele Bulivou.
\newblock Information architecture: Using best merge method, category validity, and multidimensional scaling for open card sort data analysis.
\newblock \emph{International Journal of Human–Computer Interaction}, 40\penalty0 (2):\penalty0 203--223, 2022.
\newblock URL \url{https://doi.org/10.1080/10447318.2022.2112077}.

\bibitem[Katsanos et~al.(2023)Katsanos, Christoforidis, and Demertzi]{katsanos2023}
Christos Katsanos, Vasileios Christoforidis, and Christina Demertzi.
\newblock Task-based open card sorting: Towards a new method to produce usable information architectures.
\newblock In Hirohiko Mori and Yumi Asahi, editors, \emph{Human Interface and the Management of Information}, pages 68--80, Cham, 2023. Springer Nature Switzerland.
\newblock ISBN 978-3-031-35132-7.
\newblock URL \url{https://doi.org/10.1007/978-3-031-35132-7_5}.

\bibitem[Katsanos et~al.(2022)Katsanos, Zafeiriou, and Liapis]{katsanos2022}
Christos Katsanos, Georgia Zafeiriou, and Alexandros Liapis.
\newblock Effect of self-efficacy on open card sorts for websites.
\newblock In Sakae Yamamoto and Hirohiko Mori, editors, \emph{Human Interface and the Management of Information: Visual and Information Design}, pages 75--87, Cham, 2022. Springer International Publishing.
\newblock ISBN 978-3-031-06424-1.
\newblock URL \url{https://doi.org/10.1007/978-3-031-06424-1_7}.

\bibitem[Best et~al.(2022)Best, Badham, McConnell, and and]{best2021}
Paul Best, Jennifer Badham, Tracey McConnell, and Ruth F~Hunter and.
\newblock Participatory theme elicitation: open card sorting for user led qualitative data analysis.
\newblock \emph{International Journal of Social Research Methodology}, 25\penalty0 (2):\penalty0 213--231, 2022.
\newblock URL \url{https://doi.org/10.1080/13645579.2021.1876616}.

\bibitem[Jiang et~al.(2021{\natexlab{a}})Jiang, Wang, Lin, and Shangguan]{jiang2021a}
Tingting Jiang, Ying Wang, Tianqianjin Lin, and Lina Shangguan.
\newblock Evaluating chinese government wechat official accounts in public service delivery: A user-centered approach.
\newblock \emph{Government Information Quarterly}, 38\penalty0 (1):\penalty0 101548, 2021{\natexlab{a}}.
\newblock ISSN 0740-624X.
\newblock URL \url{https://doi.org/10.1016/j.giq.2020.101548}.

\bibitem[Mac{\'i}as and Cul{\'e}n(2021)]{macias2021}
Jos{\'e}~A. Mac{\'i}as and Alma~L. Cul{\'e}n.
\newblock Enhancing decision-making in user-centered web development: a methodology for card-sorting analysis.
\newblock \emph{World Wide Web}, 24\penalty0 (6):\penalty0 2099--2137, Nov 2021.
\newblock ISSN 1573-1413.
\newblock URL \url{https://doi.org/10.1007/s11280-021-00950-y}.

\bibitem[Naranjo-Rojas et~al.(2022)Naranjo-Rojas, Ángel Perula-de Torres, and Molina-Recio]{naranjorojas2022}
Anisbed Naranjo-Rojas, Luis Ángel Perula-de Torres, and Guillermo Molina-Recio.
\newblock Patients, caregivers, and healthcare professionals' needs when designing the content of a mobile application for the clinical monitoring of patients with chronic obstructive pulmonary disease and home oxygen therapy: A user-centered design.
\newblock \emph{Internet Interventions}, 29:\penalty0 100552, 2022.
\newblock ISSN 2214-7829.
\newblock URL \url{https://doi.org/10.1016/j.invent.2022.100552}.

\bibitem[Fauth et~al.(2024)Fauth, Bloch, Noardo, Nisbet, Kaiser, {Nørkjær Gade}, and Tekavec]{fauth2024}
Judith Fauth, Tanya Bloch, Francesca Noardo, Nicholas Nisbet, Stefanie-Brigitte Kaiser, Peter {Nørkjær Gade}, and Jernej Tekavec.
\newblock Taxonomy for building permit system - organizing knowledge for building permit digitalization.
\newblock \emph{Advanced Engineering Informatics}, 59:\penalty0 102312, 2024.
\newblock ISSN 1474-0346.
\newblock URL \url{https://doi.org/10.1016/j.aei.2023.102312}.

\bibitem[Feine et~al.(2019)Feine, Gnewuch, Morana, and Maedche]{feine2019}
Jasper Feine, Ulrich Gnewuch, Stefan Morana, and Alexander Maedche.
\newblock A taxonomy of social cues for conversational agents.
\newblock \emph{International Journal of Human-Computer Studies}, 132:\penalty0 138--161, 2019.
\newblock ISSN 1071-5819.
\newblock URL \url{https://doi.org/10.1016/j.ijhcs.2019.07.009}.

\bibitem[Gerosa et~al.(2024)Gerosa, Trinkenreich, Steinmacher, and Sarma]{gerosa2024}
Marco Gerosa, Bianca Trinkenreich, Igor Steinmacher, and Anita Sarma.
\newblock Can ai serve as a substitute for human subjects in software engineering research?
\newblock \emph{Automated Software Engineering}, 31\penalty0 (1):\penalty0 13, Jan 2024.
\newblock ISSN 1573-7535.
\newblock URL \url{https://doi.org/10.1007/s10515-023-00409-6}.

\bibitem[Krajcovic et~al.(2025)Krajcovic, Demcak, and Kuric]{krajcovic2025}
Matus Krajcovic, Peter Demcak, and Eduard Kuric.
\newblock Is usability testing valid with prototypes where clickable hotspots are highlighted upon misclick?
\newblock \emph{Journal of Systems and Software}, 226:\penalty0 112446, 2025.
\newblock ISSN 0164-1212.
\newblock URL \url{https://doi.org/10.1016/j.jss.2025.112446}.

\bibitem[Ishaq et~al.(2021)Ishaq, Rosdi, Zin, and Abid]{ishaq2021}
Kashif Ishaq, Fadhilah Rosdi, Nor Azan~Mat Zin, and Adnan Abid.
\newblock Heuristic and think aloud method to evaluate the low fidelity prototype of game-based language learning application.
\newblock In \emph{2021 International Conference on Innovative Computing (ICIC)}, pages 1--10, 2021.
\newblock URL \url{https://doi.org/10.1109/ICIC53490.2021.9693022}.

\bibitem[Rosenfeld et~al.(2015)Rosenfeld, Morville, and Arango]{rosenfeld2015}
L.~Rosenfeld, P.~Morville, and J.~Arango.
\newblock \emph{Information Architecture: For the Web and Beyond}.
\newblock O'Reilly Media, 2015.
\newblock ISBN 9781491913543.

\bibitem[Schall(2014)]{schall2014}
Andrew Schall.
\newblock 6 - information architecture and web navigation.
\newblock In Jennifer {Romano Bergstrom} and Andrew~Jonathan Schall, editors, \emph{Eye Tracking in User Experience Design}, pages 139--162. Morgan Kaufmann, Boston, 2014.
\newblock ISBN 978-0-12-408138-3.
\newblock URL \url{https://doi.org/10.1016/B978-0-12-408138-3.00006-6}.

\bibitem[Ntouvaleti and Katsanos(2022)]{ntouvaleti2022}
Maria Ntouvaleti and Christos Katsanos.
\newblock Validity of the open card sorting method for producing website information structures.
\newblock In \emph{Extended Abstracts of the 2022 CHI Conference on Human Factors in Computing Systems}, CHI EA '22, pages 1--7, New York, NY, USA, 2022. Association for Computing Machinery.
\newblock ISBN 9781450391566.
\newblock URL \url{https://doi.org/10.1145/3491101.3519734}.

\bibitem[Thomas and Johnson(2013)]{thomas2013}
Robert~L. Thomas and Ian Johnson.
\newblock Merging methodologies: Combining individual and group card sorting.
\newblock In Aaron Marcus, editor, \emph{Design, User Experience, and Usability. Design Philosophy, Methods, and Tools}, pages 417--426, Berlin, Heidelberg, 2013. Springer Berlin Heidelberg.
\newblock ISBN 978-3-642-39229-0.
\newblock URL \url{https://doi.org/10.1007/978-3-642-39229-0_45}.

\bibitem[Lewis and Hepburn(2010)]{lewis2009}
Krystal~M. Lewis and Peter Hepburn.
\newblock Open card sorting and factor analysis: a usability case study.
\newblock \emph{The Electronic Library}, 28\penalty0 (3):\penalty0 401--416, Jan 2010.
\newblock ISSN 0264-0473.
\newblock URL \url{10.1108/02640471011051981}.

\bibitem[Wentzel et~al.(2016)Wentzel, Beerlage~de Jong, and van~der Geest]{wentzel2016}
Jobke Wentzel, Nienke Beerlage~de Jong, and Thea van~der Geest.
\newblock Redesign based on card sorting: How universally applicable are card sort results?
\newblock In Vincent~G. Duffy, editor, \emph{Digital Human Modeling: Applications in Health, Safety, Ergonomics and Risk Management}, pages 381--388, Cham, 2016. Springer International Publishing.
\newblock ISBN 978-3-319-40247-5.
\newblock URL \url{https://doi.org/10.1007/978-3-319-40247-5_38}.

\bibitem[Schmettow and Sommer(2016)]{schmettow2016}
Martin Schmettow and Jan Sommer.
\newblock Linking card sorting to browsing performance--are congruent municipal websites more efficient to use?
\newblock \emph{Behaviour \& information technology}, 35\penalty0 (6):\penalty0 452--470, 2016.
\newblock URL \url{https://doi.org/10.1080/0144929X.2016.1157207}.

\bibitem[Kuric et~al.(2025{\natexlab{a}})Kuric, Demcak, and Krajcovic]{kuric2025treetest}
Eduard Kuric, Peter Demcak, and Matus Krajcovic.
\newblock Validation of information architecture: Cross-methodological comparison of tree testing variants and prototype user testing.
\newblock \emph{Information and Software Technology}, 183:\penalty0 107740, 2025{\natexlab{a}}.
\newblock ISSN 0950-5849.
\newblock URL \url{https://doi.org/10.1016/j.infsof.2025.107740}.

\bibitem[de~Jesús~Álvarez Robles et~al.(2019)de~Jesús~Álvarez Robles, Álvarez Rodríguez, Benítez-Guerrero, and Rusu]{robles2019}
Teresita de~Jesús~Álvarez Robles, Francisco~Javier Álvarez Rodríguez, Edgard Benítez-Guerrero, and Cristian Rusu.
\newblock Adapting card sorting for blind people: Evaluation of the interaction design in talkback.
\newblock \emph{Computer Standards \& Interfaces}, 66:\penalty0 103356, 2019.
\newblock ISSN 0920-5489.
\newblock URL \url{https://doi.org/10.1016/j.csi.2019.103356}.

\bibitem[Greifeneder and Bressel(2022)]{greifeneder2022}
Elke Greifeneder and Paulina Bressel.
\newblock Hybrid digital card sorting: New research technique or mere variant?
\newblock In Malte Smits, editor, \emph{Information for a Better World: Shaping the Global Future}, pages 50--67, Cham, 2022. Springer International Publishing.
\newblock ISBN 978-3-030-96960-8.
\newblock URL \url{https://doi.org/10.1007/978-3-030-96960-8_4}.

\bibitem[Tchivi et~al.(2025)Tchivi, Sharma, and Paea]{tchivi2025}
Elinda Tchivi, Bibhya Sharma, and Sione Paea.
\newblock A systematic review of the comparison of different types of card sorting.
\newblock \emph{IEEE Access}, 13:\penalty0 52334--52352, 2025.
\newblock URL \url{https://doi.org/10.1109/ACCESS.2025.3552949}.

\bibitem[Lantz et~al.(2019)Lantz, Keeley, Roberts, Medina-Mora, Sharan, and Reed]{lantz2019}
Ethan Lantz, Jared~W. Keeley, Michael~C. Roberts, Maria~Elena Medina-Mora, Pratap Sharan, and Geoffrey~M. Reed.
\newblock Card sorting data collection methodology: How many participants is most efficient?
\newblock \emph{Journal of Classification}, 36\penalty0 (3):\penalty0 649--658, Oct 2019.
\newblock ISSN 1432-1343.
\newblock URL \url{https://doi.org/10.1007/s00357-018-9292-8}.

\bibitem[Bussolon(2009)]{bussolon2008}
Stefano Bussolon.
\newblock Card sorting, category validity, and contextual navigation.
\newblock \emph{Journal of Information Architecture}, 1\penalty0 (2), 2009.

\bibitem[Mart{\'\i}n and Mac{\'\i}as(2023)]{martin2023}
Marina Mart{\'\i}n and Jos{\'e}~A Mac{\'\i}as.
\newblock A supporting tool for enhancing user’s mental model elicitation and decision-making in user experience research.
\newblock \emph{International Journal of Human--Computer Interaction}, 39\penalty0 (1):\penalty0 183--202, 2023.
\newblock URL \url{https://doi.org/10.1080/10447318.2022.2041885}.

\bibitem[Ali et~al.(2019)Ali, Ashby, Webb, Zwitser, and Cesar]{ali2019}
Abdallah~El Ali, Liam Ashby, Andrew~M. Webb, Robert Zwitser, and Pablo Cesar.
\newblock Uncovering perceived identification accuracy of in-vehicle biometric sensing.
\newblock In \emph{Proceedings of the 11th International Conference on Automotive User Interfaces and Interactive Vehicular Applications: Adjunct Proceedings}, AutomotiveUI '19, page 327–334, New York, NY, USA, 2019. Association for Computing Machinery.
\newblock ISBN 9781450369206.
\newblock URL \url{https://doi.org/10.1145/3349263.3351506}.

\bibitem[Righi et~al.(2013)Righi, James, Beasley, Day, Fox, Gieber, Howe, and Ruby]{righi2013}
Carol Righi, Janice James, Michael Beasley, Donald~L Day, Jean~E Fox, Jennifer Gieber, Chris Howe, and Laconya Ruby.
\newblock Card sort analysis best practices.
\newblock \emph{Journal of Usability Studies}, 8\penalty0 (3):\penalty0 69--89, 2013.

\bibitem[Gabe-Thomas et~al.(2016)Gabe-Thomas, Walker, Verplanken, and Shaddick]{gabethomas2016}
Elizabeth Gabe-Thomas, Ian Walker, Bas Verplanken, and Gavin Shaddick.
\newblock Householders’ mental models of domestic energy consumption: Using a sort-and-cluster method to identify shared concepts of appliance similarity.
\newblock \emph{PLOS ONE}, 11\penalty0 (7):\penalty0 1--15, 07 2016.
\newblock URL \url{https://doi.org/10.1371/journal.pone.0158949}.

\bibitem[Robertson et~al.(2020)Robertson, Kortum, Oswald, and Acemyan]{robertson2021}
Ian Robertson, Philip Kortum, Frederick~L. Oswald, and Claudia~Ziegler Acemyan.
\newblock Novices perform like experts on a closed card sort but not an open card sort.
\newblock \emph{Proceedings of the Human Factors and Ergonomics Society Annual Meeting}, 64\penalty0 (1):\penalty0 1249--1253, 2020.
\newblock URL \url{https://doi.org/10.1177/1071181320641297}.

\bibitem[Lu et~al.(2024{\natexlab{a}})Lu, Yang, Zhao, Zhang, and Li]{lu2024b}
Yuwen Lu, Yuewen Yang, Qinyi Zhao, Chengzhi Zhang, and Toby Jia-Jun Li.
\newblock Ai assistance for ux: A literature review through human-centered ai, 2024{\natexlab{a}}.
\newblock URL \url{https://arxiv.org/abs/2402.06089}.

\bibitem[Raees et~al.(2024)Raees, Meijerink, Lykourentzou, Khan, and Papangelis]{raees2024}
Muhammad Raees, Inge Meijerink, Ioanna Lykourentzou, Vassilis-Javed Khan, and Konstantinos Papangelis.
\newblock From explainable to interactive ai: A literature review on current trends in human-ai interaction.
\newblock \emph{International Journal of Human-Computer Studies}, 189:\penalty0 103301, 2024.
\newblock ISSN 1071-5819.
\newblock URL \url{https://doi.org/10.1016/j.ijhcs.2024.103301}.

\bibitem[Capel and Brereton(2023)]{capel2023}
Tara Capel and Margot Brereton.
\newblock What is human-centered about human-centered ai? a map of the research landscape.
\newblock In \emph{Proceedings of the 2023 CHI Conference on Human Factors in Computing Systems}, CHI '23, pages 1--23, New York, NY, USA, 2023. Association for Computing Machinery.
\newblock ISBN 9781450394215.
\newblock URL \url{https://doi.org/10.1145/3544548.3580959}.

\bibitem[Aitim and Abdulla(2024)]{aitim2024}
Aigerim Aitim and Muslima Abdulla.
\newblock Data processing and analysing techniques in ux research.
\newblock \emph{Procedia Computer Science}, 251:\penalty0 591--596, 2024.
\newblock ISSN 1877-0509.
\newblock URL \url{https://doi.org/10.1016/j.procs.2024.11.154}.
\newblock 15th International Conference on Emerging Ubiquitous Systems and Pervasive Networks / 14th International Conference on Current and Future Trends of Information and Communication Technologies in Healthcare EUSPN/ICTH 2024.

\bibitem[Borlinghaus and Huber(2021)]{borlinghaus2021}
Parzival Borlinghaus and Stephan Huber.
\newblock Comparing apples and oranges: Human and computer clustered affinity diagrams under the microscope.
\newblock In \emph{Proceedings of the 26th International Conference on Intelligent User Interfaces}, IUI '21, page 413–422, New York, NY, USA, 2021. Association for Computing Machinery.
\newblock ISBN 9781450380171.
\newblock URL \url{https://doi.org/10.1145/3397481.3450674}.

\bibitem[Jiang et~al.(2021{\natexlab{b}})Jiang, Wade, Fiesler, and Brubaker]{jiang2021b}
Jialun~Aaron Jiang, Kandrea Wade, Casey Fiesler, and Jed~R. Brubaker.
\newblock Supporting serendipity: Opportunities and challenges for human-ai collaboration in qualitative analysis.
\newblock \emph{Proc. ACM Hum.-Comput. Interact.}, 5\penalty0 (CSCW1), April 2021{\natexlab{b}}.
\newblock URL \url{https://doi.org/10.1145/3449168}.

\bibitem[Guzman and Maalej(2014)]{guzman2014}
Emitza Guzman and Walid Maalej.
\newblock How do users like this feature? a fine grained sentiment analysis of app reviews.
\newblock In \emph{2014 IEEE 22nd International Requirements Engineering Conference (RE)}, pages 153--162, 2014.
\newblock URL \url{https://doi.org/10.1109/RE.2014.6912257}.

\bibitem[Jang and Park(2022)]{jang2022}
Yeonju Jang and Eunil Park.
\newblock Satisfied or not: user experience of mobile augmented reality in using natural language processing techniques on review comments.
\newblock \emph{Virtual Reality}, 26\penalty0 (3):\penalty0 839--848, Sep 2022.
\newblock ISSN 1434-9957.
\newblock URL \url{https://doi.org/10.1007/s10055-021-00599-y}.

\bibitem[Schuller et~al.(2024)Schuller, Janssen, Blumenr\"{o}ther, Probst, Schmidt, and Kumar]{schuller2024}
Andreas Schuller, Doris Janssen, Julian Blumenr\"{o}ther, Theresa~Maria Probst, Michael Schmidt, and Chandan Kumar.
\newblock Generating personas using llms and assessing their viability.
\newblock In \emph{Extended Abstracts of the CHI Conference on Human Factors in Computing Systems}, CHI EA '24, pages 1--7, New York, NY, USA, 2024. Association for Computing Machinery.
\newblock ISBN 9798400703317.
\newblock URL \url{https://doi.org/10.1145/3613905.3650860}.

\bibitem[Gessinger et~al.(2025)Gessinger, Seaborn, Steeds, and Cowan]{gessinger2025}
Iona Gessinger, Katie Seaborn, Madeleine Steeds, and Benjamin~R. Cowan.
\newblock Chatgpt and me: First-time and experienced users’ perceptions of chatgpt’s communicative ability as a dialogue partner.
\newblock \emph{International Journal of Human-Computer Studies}, 194:\penalty0 103400, 2025.
\newblock ISSN 1071-5819.
\newblock URL \url{https://doi.org/10.1016/j.ijhcs.2024.103400}.

\bibitem[Xiao et~al.(2020)Xiao, Zhou, Liao, Mark, Chi, Chen, and Yang]{xiao2020}
Ziang Xiao, Michelle~X. Zhou, Q.~Vera Liao, Gloria Mark, Changyan Chi, Wenxi Chen, and Huahai Yang.
\newblock Tell me about yourself: Using an ai-powered chatbot to conduct conversational surveys with open-ended questions.
\newblock \emph{ACM Trans. Comput.-Hum. Interact.}, 27\penalty0 (3), June 2020.
\newblock ISSN 1073-0516.
\newblock URL \url{https://doi.org/10.1145/3381804}.

\bibitem[Kuric et~al.(2024)Kuric, Demcak, and Krajcovic]{kuric2024gpt}
Eduard Kuric, Peter Demcak, and Matus Krajcovic.
\newblock Unmoderated usability studies evolved: Can gpt ask useful follow-up questions?
\newblock \emph{International Journal of Human–Computer Interaction}, 0\penalty0 (0):\penalty0 1--18, 2024.
\newblock URL \url{https://doi.org/10.1080/10447318.2024.2427978}.

\bibitem[Rothschild et~al.(2024)Rothschild, Brand, Schroeder, and Wang]{rothschild2024}
David~M Rothschild, James Brand, Hope Schroeder, and Jenny Wang.
\newblock Opportunities and risks of llms in survey research.
\newblock \emph{Available at SSRN}, 2024.

\bibitem[Zavod et~al.(2002)Zavod, Rickert, Brown, and Mutual]{zavod2002}
Merrill~J. Zavod, Donald~E. Rickert, Steven~H. Brown, and State~Farm Mutual.
\newblock The automated card-sort as an interface design tool: A comparison of products.
\newblock \emph{Proceedings of the Human Factors and Ergonomics Society Annual Meeting}, 46\penalty0 (5):\penalty0 646--650, 2002.
\newblock URL \url{https://doi.org/10.1177/154193120204600510}.

\bibitem[Sauro et~al.(2024)Sauro, Schiavone, and Lewis]{sauro2024}
Jeff Sauro, Will Schiavone, and Jim Lewis.
\newblock Comparing chatgpt to card sorting results, 2024.
\newblock URL \url{https://measuringu.com/comparing-chatgpt-to-card-sorting-results/}.

\bibitem[H\"{a}m\"{a}l\"{a}inen et~al.(2023)H\"{a}m\"{a}l\"{a}inen, Tavast, and Kunnari]{hamalainen2023}
Perttu H\"{a}m\"{a}l\"{a}inen, Mikke Tavast, and Anton Kunnari.
\newblock Evaluating large language models in generating synthetic hci research data: a case study.
\newblock In \emph{Proceedings of the 2023 CHI Conference on Human Factors in Computing Systems}, CHI '23, pages 1--19, New York, NY, USA, 2023. Association for Computing Machinery.
\newblock ISBN 9781450394215.
\newblock URL \url{https://doi.org/10.1145/3544548.3580688}.

\bibitem[H\"{a}m\"{a}l\"{a}inen et~al.(2022)H\"{a}m\"{a}l\"{a}inen, Tavast, and Kunnari]{hamalainen2022}
Perttu H\"{a}m\"{a}l\"{a}inen, Mikke Tavast, and Anton Kunnari.
\newblock Neural language models as what if? -engines for hci research.
\newblock In \emph{Companion Proceedings of the 27th International Conference on Intelligent User Interfaces}, IUI '22 Companion, page 77–80, New York, NY, USA, 2022. Association for Computing Machinery.
\newblock ISBN 9781450391450.
\newblock URL \url{https://doi.org/10.1145/3490100.3516458}.

\bibitem[Gu et~al.(2025)Gu, Chandrasegaran, and Lloyd]{gu2024}
(Eric)~Heng Gu, Senthil Chandrasegaran, and Peter Lloyd.
\newblock Synthetic users: insights from designers’ interactions with persona-based chatbots.
\newblock \emph{Artificial Intelligence for Engineering Design, Analysis and Manufacturing}, 39:\penalty0 e2, 2025.
\newblock URL \url{https://doi.org/10.1017/S0890060424000283}.

\bibitem[Kuric et~al.(2025{\natexlab{b}})Kuric, Demcak, Krajcovic, and Lang]{kuric2025slr}
Eduard Kuric, Peter Demcak, Matus Krajcovic, and Jan Lang.
\newblock Systematic literature review of automation and artificial intelligence in usability issue detection, 2025{\natexlab{b}}.
\newblock URL \url{https://arxiv.org/abs/2504.01415}.

\bibitem[Xiang et~al.(2024)Xiang, Zhu, Lou, Chen, Pan, Jin, Chen, and Sun]{xiang2024}
Wei Xiang, Hanfei Zhu, Suqi Lou, Xinli Chen, Zhenghua Pan, Yuping Jin, Shi Chen, and Lingyun Sun.
\newblock Simuser: Generating usability feedback by simulating various users interacting with mobile applications.
\newblock In \emph{Proceedings of the 2024 CHI Conference on Human Factors in Computing Systems}, CHI '24, pages 1--17, New York, NY, USA, 2024. Association for Computing Machinery.
\newblock ISBN 9798400703300.
\newblock URL \url{https://doi.org/10.1145/3613904.3642481}.

\bibitem[Kim and Lee(2024)]{kim2024}
Junsol Kim and Byungkyu Lee.
\newblock Ai-augmented surveys: Leveraging large language models and surveys for opinion prediction, 2024.
\newblock URL \url{https://arxiv.org/abs/2305.09620}.

\bibitem[Sanders et~al.(2023)Sanders, Ulinich, and Schneier]{sanders2023}
Nathan~E. Sanders, Alex Ulinich, and Bruce Schneier.
\newblock Demonstrations of the potential of ai-based political issue polling, 2023.
\newblock URL \url{https://arxiv.org/abs/2307.04781}.

\bibitem[Li et~al.(2025)Li, Zhang, Zhang, Zhang, Liu, Yao, Xu, Zheng, Wang, Chen, Zhang, Yin, Dong, Li, Bi, Mei, Fang, Guo, Song, and Liu]{li2025}
Zhong-Zhi Li, Duzhen Zhang, Ming-Liang Zhang, Jiaxin Zhang, Zengyan Liu, Yuxuan Yao, Haotian Xu, Junhao Zheng, Pei-Jie Wang, Xiuyi Chen, Yingying Zhang, Fei Yin, Jiahua Dong, Zhiwei Li, Bao-Long Bi, Ling-Rui Mei, Junfeng Fang, Zhijiang Guo, Le~Song, and Cheng-Lin Liu.
\newblock From system 1 to system 2: A survey of reasoning large language models, 2025.
\newblock URL \url{https://arxiv.org/abs/2502.17419}.

\bibitem[Esposito et~al.(2024)Esposito, Desolda, and Lanzilotti]{esposito2024}
Andrea Esposito, Giuseppe Desolda, and Rosa Lanzilotti.
\newblock The fine line between automation and augmentation in website usability evaluation.
\newblock \emph{Scientific Reports}, 14\penalty0 (1):\penalty0 10129, May 2024.
\newblock ISSN 2045-2322.
\newblock URL \url{https://doi.org/10.1038/s41598-024-59616-0}.

\bibitem[OpenAI(2024)]{openai2024}
OpenAI.
\newblock Gpt-4o system card, 2024.
\newblock URL \url{https://arxiv.org/abs/2410.21276}.

\bibitem[Google(2024)]{google2023}
Google.
\newblock Gemini: A family of highly capable multimodal models, 2024.
\newblock URL \url{https://arxiv.org/abs/2312.11805}.

\bibitem[Lu et~al.(2024{\natexlab{b}})Lu, Aleta, Du, Shi, and Moreno]{lu2024}
Yikang Lu, Alberto Aleta, Chunpeng Du, Lei Shi, and Yamir Moreno.
\newblock Llms and generative agent-based models for complex systems research.
\newblock \emph{Physics of Life Reviews}, 51:\penalty0 283--293, 2024{\natexlab{b}}.
\newblock ISSN 1571-0645.
\newblock URL \url{https://doi.org/10.1016/j.plrev.2024.10.013}.

\bibitem[Lippert et~al.(2024)Lippert, Dreber, Johannesson, Tierney, Cyrus-Lai, Uhlmann, null, and Pfeiffer]{lippert2024}
Steffen Lippert, Anna Dreber, Magnus Johannesson, Warren Tierney, Wilson Cyrus-Lai, Eric~Luis Uhlmann, null null, and Thomas Pfeiffer.
\newblock Can large language models help predict results from a complex behavioural science study?
\newblock \emph{Royal Society Open Science}, 11\penalty0 (9):\penalty0 240682, 2024.
\newblock URL \url{https://doi.org/10.1098/rsos.240682}.

\bibitem[Wang et~al.(2023)Wang, Li, Yin, Wu, and Liu]{wang2023}
Xuena Wang, Xueting Li, Zi~Yin, Yue Wu, and Jia Liu.
\newblock Emotional intelligence of large language models.
\newblock \emph{Journal of Pacific Rim Psychology}, 17:\penalty0 18344909231213958, 2023.
\newblock URL \url{https://doi.org/10.1177/18344909231213958}.

\bibitem[Chang et~al.(2024)Chang, Wang, Wang, Wu, Yang, Zhu, Chen, Yi, Wang, Wang, Ye, Zhang, Chang, Yu, Yang, and Xie]{chang2024}
Yupeng Chang, Xu~Wang, Jindong Wang, Yuan Wu, Linyi Yang, Kaijie Zhu, Hao Chen, Xiaoyuan Yi, Cunxiang Wang, Yidong Wang, Wei Ye, Yue Zhang, Yi~Chang, Philip~S. Yu, Qiang Yang, and Xing Xie.
\newblock A survey on evaluation of large language models.
\newblock \emph{ACM Trans. Intell. Syst. Technol.}, 15\penalty0 (3), March 2024.
\newblock ISSN 2157-6904.
\newblock URL \url{https://doi.org/10.1145/3641289}.

\bibitem[Marvin et~al.(2024)Marvin, Hellen, Jjingo, and Nakatumba-Nabende]{marvin2024}
Ggaliwango Marvin, Nakayiza Hellen, Daudi Jjingo, and Joyce Nakatumba-Nabende.
\newblock Prompt engineering in large language models.
\newblock In I.~Jeena Jacob, Selwyn Piramuthu, and Przemyslaw Falkowski-Gilski, editors, \emph{Data Intelligence and Cognitive Informatics}, pages 387--402, Singapore, 2024. Springer Nature Singapore.
\newblock ISBN 978-981-99-7962-2.

\bibitem[White et~al.(2023)White, Fu, Hays, Sandborn, Olea, Gilbert, Elnashar, Spencer-Smith, and Schmidt]{white2023}
Jules White, Quchen Fu, Sam Hays, Michael Sandborn, Carlos Olea, Henry Gilbert, Ashraf Elnashar, Jesse Spencer-Smith, and Douglas~C Schmidt.
\newblock A prompt pattern catalog to enhance prompt engineering with chatgpt, 2023.
\newblock URL \url{https://doi.org/10.48550/arXiv.2302.11382}.

\bibitem[Zhao et~al.(2025)Zhao, Zhou, Li, Tang, Wang, Hou, Min, Zhang, Zhang, Dong, Du, Yang, Chen, Chen, Jiang, Ren, Li, Tang, Liu, Liu, Nie, and Wen]{zhao2023}
Wayne~Xin Zhao, Kun Zhou, Junyi Li, Tianyi Tang, Xiaolei Wang, Yupeng Hou, Yingqian Min, Beichen Zhang, Junjie Zhang, Zican Dong, Yifan Du, Chen Yang, Yushuo Chen, Zhipeng Chen, Jinhao Jiang, Ruiyang Ren, Yifan Li, Xinyu Tang, Zikang Liu, Peiyu Liu, Jian-Yun Nie, and Ji-Rong Wen.
\newblock A survey of large language models, 2025.
\newblock URL \url{https://arxiv.org/abs/2303.18223}.

\bibitem[OpenAI(2025)]{openai2025}
OpenAI.
\newblock Text generation and prompting, 2025.
\newblock URL \url{https://platform.openai.com/docs/guides/text}.
\newblock Accessed: 2025-04-22.

\bibitem[Satopaa et~al.(2011)Satopaa, Albrecht, Irwin, and Raghavan]{satopaa2011}
Ville Satopaa, Jeannie Albrecht, David Irwin, and Barath Raghavan.
\newblock Finding a "kneedle" in a haystack: Detecting knee points in system behavior.
\newblock In \emph{2011 31st International Conference on Distributed Computing Systems Workshops}, pages 166--171, 2011.
\newblock URL \url{https://doi.org/10.1109/ICDCSW.2011.20}.

\bibitem[Zheng et~al.(2020)Zheng, Zhu, Wen, Zhu, Yu, and Gan]{zheng2020}
Wei Zheng, Xiaofeng Zhu, Guoqiu Wen, Yonghua Zhu, Hao Yu, and Jiangzhang Gan.
\newblock Unsupervised feature selection by self-paced learning regularization.
\newblock \emph{Pattern Recognition Letters}, 132:\penalty0 4--11, 2020.
\newblock ISSN 0167-8655.
\newblock URL \url{https://doi.org/10.1016/j.patrec.2018.06.029}.
\newblock Multiple-Task Learning for Big Data (MTL4BD).

\bibitem[Cady(2024)]{cady2024}
Field Cady.
\newblock \emph{The data science handbook}.
\newblock John Wiley \& Sons, 2024.

\bibitem[Nawaz(2012)]{nawaz2012}
Ather Nawaz.
\newblock A comparison of card-sorting analysis methods.
\newblock In \emph{APCHI'12. Proceedings of the 10th Asia Pacific Conference on Computer-Human Interaction}, pages 583--592. Association for Computing Machinery, 2012.

\bibitem[Stiennon et~al.(2020)Stiennon, Ouyang, Wu, Ziegler, Lowe, Voss, Radford, Amodei, and Christiano]{stiennon2020}
Nisan Stiennon, Long Ouyang, Jeffrey Wu, Daniel Ziegler, Ryan Lowe, Chelsea Voss, Alec Radford, Dario Amodei, and Paul~F Christiano.
\newblock Learning to summarize with human feedback.
\newblock In H.~Larochelle, M.~Ranzato, R.~Hadsell, M.F. Balcan, and H.~Lin, editors, \emph{Advances in Neural Information Processing Systems}, volume~33, pages 3008--3021. Curran Associates, Inc., 2020.
\newblock URL \url{https://proceedings.neurips.cc/paper_files/paper/2020/file/1f89885d556929e98d3ef9b86448f951-Paper.pdf}.

\end{thebibliography}

\end{document}